\documentclass[11pt,a4paper]{article}
 \usepackage[latin1]{inputenc}
 \usepackage[UKenglish]{babel}
 \usepackage{myway}
 \usepackage{graphicx}
 \usepackage{epstopdf}

\hypersetup{pdftitle={Fluxes and Chiral Matter}, 
pdfauthor={Sven Krause, Christoph Mayrhofer, Timo Weigand}, 
pdfsubject={String theory model building, F-theory},
bookmarksopen, bookmarksnumbered, bookmarksopenlevel=2}

\newcommand{\beq}{\begin{equation}} 
 \newcommand{\eeq}{\end{equation}}
\newcommand{\bal}{\begin{aligned}}  
 \newcommand{\eal}{\end{aligned}}
\newcommand{\bea}{\begin{eqnarray}} 
 \newcommand{\eea}{\end{eqnarray}}
 \def\ov{\overline}

\newcommand{\bbP}{\mathbb{P}}

\newcommand{\tw}{\text{w}}
\newcommand{\tW}{\text{W}}

\newcommand{\aK}{\bar{K}}
\newcommand{\pl}[1]{\bbP^1_{\,#1}}

\newcommand{\executeiffilenewer}[3]{%
 \ifnum\pdfstrcmp{\pdffilemoddate{#1}}%
 {\pdffilemoddate{#2}}>0%
 {\immediate\write18{#3}}\fi%
}

\title{$G_4$ flux, chiral matter and singularity resolution in F-theory compactifications }

\preprint{}

\author[1]%
{Sven Krause,%\note[1]{\href{mailto:S.Krause@ThPhys.Uni-Heidelberg.de}{S.Krause@ThPhys.Uni-Heidelberg.de}}%
}
\author[1]%
{Christoph Mayrhofer%\note[2]{\href{mailto:C.Mayrhofer@ThPhys.Uni-Heidelberg.de}{C.Mayrhofer@ThPhys.Uni-Heidelberg.de}}%
}
\author[1,2]%
{and Timo Weigand%\note[3]{\href{mailto:T.Weigand@ThPhys.Uni-Heidelberg.de}{T.Weigand@ThPhys.Uni-Heidelberg.de}}%
}
\affiliation[1]{Institut f\"ur Theoretische Physik, Universit\"at Heidelberg, Philosophenweg 19, D-69120 Heidelberg\vspace{0.1cm}}
\affiliation[2]{Kavli Institute for Theoretical Physics China, CAS, Beijing 100190, China\vspace{0.1cm}}

\emailAdd{S.Krause@ThPhys.Uni-Heidelberg.de} 
\emailAdd{C.Mayrhofer@ThPhys.Uni-Heidelberg.de}
\emailAdd{T.Weigand@ThPhys.Uni-Heidelberg.de}

\abstract{ 

We construct a set of chirality inducing $G_4$-fluxes in global F-theory compactifications on Calabi-Yau four-folds.
Special emphasis is put on models with gauge group $SU(5) \times U(1)_X$ relevant in the context of F-theory GUT model building, which are described in terms of a  $U(1)$-restricted Tate model.
A $G_4$-flux arises in a manner completely analogous to the $U(1)_X$ gauge potential.
We describe in detail the resolution by blow-up of the various singularities responsible for the $U(1)_X$ factor and the standard $SU(5)$ gauge group and match the result with techniques applied in the context of toric geometry.
This provides an explicit identification of the structure of the resolved fibre over the matter curves and over the enhancement points relevant for Yukawa couplings.
We compute the flux-induced chiral index both of $SU(5)$ charged matter and of $SU(5)$ singlets charged only under $U(1)_X$ localised on curves which are not contained in the $SU(5)$ locus.
We furthermore discuss global consistency conditions such as D3-tadpole cancellation, D-term supersymmetry  and Freed-Witten quantisation.
The $U(1)_X$ gauge flux is a global extension of a class of split spectral cover bundles. 
It constitutes an essential ingredient in the construction of globally defined F-theory compactifications with chiral matter. 
We exemplify this in a three-generation $SU(5) \times U(1)_X$ model whose flux satisfies all of the above global consistency conditions. 
We also extend our results to chiral fluxes in models without $U(1)$ restriction.

}
\begin{document}

\maketitle

\section{Introduction}

F-theory~\cite{Vafa:1996xn} provides an elegant framework to study a very broad class of string vacua.  Its power and its beauty are rooted in the geometrisation of the back-reaction of physical objects, here seven-branes of Type IIB string theory,  on the ambient space. This is achieved by means of a non-trivial fibration of an auxiliary elliptic curve over the physical space-time; its complex structure represents the varying axio-dilaton sourced by the seven-branes. The holomorphic nature of the relevant geometric data --- seven-branes wrap divisors of the base of the fibration upon compactification to four dimensions --- makes the study of the associated string vacua amenable to techniques of algebraic geometry. These geometric methods give us insights into systems beyond the perturbative realm such as mutually non-local [p,q]-seven-branes.
The resulting marriage between the concept of brane localised gauge degrees of freedom and the appearance of exceptional gauge groups is largely responsible for the revived recent interest, triggered by~\cite{Donagi:2008ca,Beasley:2008dc,Beasley:2008kw,Donagi:2008kj,Hayashi:2008ba}, in F-theory also from a phenomenological perspective (see~\cite{Denef:2008wq,Heckman:2010bq,Weigand:2010wm} for reviews on F-theory and its recent applications).

Motivated by the prospects of local F-theory model building in the context of GUT phenomenology, a great deal of recent effort has gone into the construction of globally consistent four-dimensional F-theory vacua. From the start, it has been clear that the key to the construction of such vacua and to understanding their properties is having a handle on the singularity structure of elliptic four-folds. This is because the non-abelian gauge groups, the matter spectrum and the Yukawa interactions of a model are in one-to-one correspondence with the singularities in the fibre of the Calabi-Yau four-fold over loci of, respectively, complex co-dimension one, two and three on the base (see~\cite{Bershadsky:1996nh} for a description of the relevant Tate algorithm and~\cite{Morrison:2011mb,Katz:2011qp} for more recent extensions thereof). 
In order to make sense of the four-dimensional effective action via dimensional reduction of the dual M-theory, discussed in detail in~\cite{Grimm:2010ks}, it is necessary to work not with this singular four-fold $Y_4$, but rather with a resolved Calabi-Yau $\hat Y_4$. Mathematically, the singular points in the fibre are replaced by a collection of ${\mathbb P^1}s$ whose intersection structure reproduces the Dynkin diagram of the simple group  associated with the singularity. Physically, resolving this singularity corresponds to moving in the Coulomb branch of the non-abelian gauge groups in the dual M-theory. In the F-theory limit of vanishing fibre volume, the  resolved space $\hat Y_4$ and the singular $Y_4$ are indistinguishable. However, it is in terms of the smooth and well-defined $\hat Y_4$ that all computations are performed.

\subsection{Singular elliptic fibrations and their resolutions}

The  techniques for resolution of singular elliptic fibrations were applied to F-theory soon after its discovery, starting mainly in compactifications to six dimensions.
Most notably, using the powerful tools of toric geometry, an efficient algorithm was developed to completely resolve singular Calabi-Yau three-folds that are hypersurfaces of toric spaces~\cite{Candelas:1996su,Candelas:1997eh}. 
In the context of F-theory GUT model building the first complete resolutions of Calabi-Yau four-folds with $SU(5)$ gauge group, as required in the spirit of~\cite{Donagi:2008ca,Beasley:2008dc,Beasley:2008kw,Donagi:2008kj}, were constructed in~\cite{Blumenhagen:2009yv,Grimm:2009yu}. This was done likewise in the framework of toric geometry, generalising the methods of~\cite{Candelas:1996su,Candelas:1997eh} to four-folds constructed as complete intersections of toric ambient spaces. As demonstrated in~\cite{Chen:2010ts,Knapp:2011wk}, the efficiency of the toric approach allows for a systematic construction and study of a  large set of four-dimensional F-theory GUT vacua which, in particular, comprises the full four-fold associated with the base space constructed previously in~\cite{Marsano:2009ym}, see also~\cite{mayrhofer:diss}.  
It is important to stress that the toric resolution automatically takes care not only of the co-dimension one singularities, corresponding to seven-branes, but also of the higher co-dimensional singularities along matter curves and Yukawa points. What the construction provides is the blow-up of the singularities over the divisors where they appeared. Thereby, the blow-up introduces a set of rk$(G)$ extra blow-up divisors fibred over the base divisor associated with gauge group $G$. This automatically resolves also the higher co-dimension singularities.
In particular, one has full computational control over the intersection properties of the resolution divisors, the complete set of Hodge numbers and important topological invariants such as the Euler characteristic or $c_2(\hat Y_4)$. These invariants enter phenomenologically relevant constraints such as  the three-brane tadpole~\cite{Sethi:1996es} or the flux quantisation condition~\cite{Witten:1996md}. 
On the other hand, the \emph{structure} of the singularity enhancements over the matter curves and Yukawa points is rather implicitly contained in the toric data, see e.g.~\cite{Intriligator:1997pq}. For practical computations, it is often desirable to have more direct access to this information.

More recently, the resolution of singular four-folds with $SU(5)$ GUT symmetries has been re-addressed in~\cite{Esole:2011sm}  using a different method corresponding to a  small resolution, as opposed to a blow-up, with special emphasis on the matter curves and Yukawa points. Indeed this analysis has confirmed the general philosophy of higher singularities of $SO(10)$ and $SU(6)$ type over matter curves in generic $SU(5)$ models as well as the appearance of $SO(12)$ and $E_6$ enhancement points. At a technical level, however, the structure especially of the $E_6$ point is more complicated than usually anticipated. There, the singularity corresponds to the non-extended $E_6$ Dynkin diagram or $T^-_{3,3,3}$~\cite{Esole:2011sm}. In the recent work of~\cite{Marsano:2011hv}, amongst other things, the consequences of these technical subtleties were analysed. The authors found that the expected structure both of matter states and, in particular, of their couplings at the enhancement points is unaffected.

\subsection{\texorpdfstring{$G_4$}{G4} fluxes from \texorpdfstring{$U(1)$}{U(1)}-restricted Tate models}

The geometry of the four-fold and its resolution, important as it is, makes  only half of the story in constructing F-theory vacua.
The second, equally crucial ingredient is $G_4$-flux. Via F/M-theory duality, $G_4$-fluxes are known to describe both what corresponds in the Type IIB limit to background flux $F_3 - \tau H_3$ and gauge flux $F$ along the seven-branes. 
Both of them
are key players in moduli stabilisation, but the latter are, in addition, indispensable in order to produce a chiral matter spectrum.

By F/M-theory duality, specifying $G_4$-fluxes amounts to choosing suitably quantised elements of $H^4(\hat Y_4)$ subject to the condition that the four-form has 'precisely one leg along the fibre'. Taking into account F-term conditions the flux must eventually be of $(2,2)$ type.\footnote{If necessary, this F-term condition will fix some of the complex structure moduli.}
The analogue of the Type IIB closed fluxes $H_3$ and $F_3$ are given by $G_4$-flux which can locally be written as a wedge product of a three-form on the base with one of the two 1-forms along the \emph{non-singular} fibre.
Gauge fluxes on the other hand have one leg along the \emph{singular} fibres or rather along the resolution of the singular fibre in $\hat Y_4$.

One type of such gauge flux that is particularly easy to understand is flux associated with the Cartan generators of the non-abelian gauge group $G$ along a divisor $W$. Such fluxes can be written as $G_4 = F \wedge \nu$, where by $\nu \in H^{1,1}(\hat Y_4)$ we denote the two-forms dual  to the resolution divisors introduced when resolving the non-abelian singularity over $W$ and $F \in H^{1,1}(W)$.
Cartan fluxes of course break the gauge group $G$. 
If we are interested in a chiral spectrum with unbroken non-abelian gauge group we thus need another type of flux.\footnote{In particular note that the hypercharge Cartan  flux used to break $SU(5) \rightarrow SU(3) \times SU(2) \times U(1)_Y$ as in~\cite{Beasley:2008kw,Donagi:2008kj}  must be chosen such as not to produce chirality as otherwise $U(1)_Y$ would become massive.}

In this paper we are interested in a type of $G_4$ gauge flux  that does produce chirality without breaking any non-abelian gauge symmetry.
We approach the construction of chirality inducing $G_4$-fluxes in the context of F-theory compactifications with explicit $U(1)$ gauge symmetries, for the following two reasons: First, we will exploit the fact that in the presence of such $U(1)$s there exists a particularly natural candidate for a special element in $H^{1,1}(\hat Y_4)$ with one leg along the fibre~\cite{Grimm:2010ez} lending itself to the construction of $G_4$-flux. Second, $U(1)$ symmetries play a prominent r{\^o}le in concrete phenomenological applications of F-theory thanks to their selection rules in the matter coupling sector (see e.g.~\cite{Hayashi:2009ge,Marsano:2009gv,Dudas:2009hu} for some early references, followed by many others). It is therefore of particular interest to construct fluxes in models with abelian gauge symmetries.

The appearance of massless $U(1)$s depends on the full global geometric data of the compactification and cannot be determined in any local approach to F-theory model building~\cite{Hayashi:2010zp},\cite{Grimm:2010ez}. 
In the context of the so-called $U(1)$-restricted Tate model~\cite{Grimm:2010ez}, an explicit construction of models with $U(1)$ gauge symmetries was given.
The idea is to start from F-theory on an elliptic Calabi-Yau four-fold $Y_4$ with no massless $U(1)$ gauge potentials and to restrict the complex structure moduli such as to unhiggs a $U(1)$ gauge symmetry. The elliptic fibre acquires a Kodaira $I_2$, or in other words an $SU(2)$ singularity over a curve $\cal C$ on the base space $B_3$. Note that this happens in co-dimension one in complex structure moduli space. The arising curve of $SU(2)$ singularities is the self-intersection locus of the $I_1$-component of the discriminant $\Delta$ of the four-fold. 
The singularity can be resolved by a blow-up procedure similar to the resolution of singularities in co-dimension one.
This gives rise to
an exceptional divisor $\{s=0\}$. Its   dual two-form leads to an extra $U(1)$ gauge potential upon expanding the M-theory three-form $C_3$ as $C_3 = A \wedge {\tw}_X$  with $\tw_X =-[S] +[Z] +c_1(B_3)$~\cite{Grimm:2010ez}.\footnote{We denote the two-form dual to the divisor in class S by the symbol $[S] \in H^{1,1}(\hat Y_4)$. Furthermore, one needs to subtract the fibre class $[Z]$ and the first Chern class of the tangent bundle of the base in order to make sure that the resulting two-form indeed has only one leg along the fibre~\cite{Grimm:2010ez}. For models involving extra non-abelian singularities, the definition of $\tw_X$ will be modified as specified in the sequel.}
Obviously, this construction provides a natural candidate for the $U(1)_X$ flux $G_4 = F \wedge \tw_X$~\cite{unpublished}. Note that this type of flux is special in that it is given by a four-form that can be written as the wedge product of two harmonic two-forms. From general arguments~\cite{Donagi:2008ca,Hayashi:2008ba},  this flux leads to a chiral index for matter states charged under $U(1)_X$ by integration of $G_4$ over the associated matter surfaces.
The reason why this conclusion could not be checked explicitly in~\cite{Grimm:2010ez} was an  insufficient understanding of the fibre structure over the matter curves as arise e.g.~in models with $SU(5)$ gauge symmetry. 
In this paper, inspired by the explicit description of fibres over the matter curves~\cite{Esole:2011sm}, we are able to explicitly compute the chiral spectrum induced by $U(1)_X$ flux.

In fact, quite recently the same type of gauge flux was independently discussed in detail in the beautiful work~\cite{Braun:2011zm}, albeit with slightly different methods. This analysis
starts with the construction of a gauge flux  which cannot be decomposed into the wedge of two-forms. In a second step the complex structure of the four-fold is  restricted  leading to a $U(1)$-restricted Tate model. Under this deformation the original gauge flux turns into flux that can be written as a wedge product. It completely agrees with the construction outlined above. 
At a technical level the construction of the flux differs slightly from ours in that the authors of~\cite{Braun:2011zm} perform a small resolution as opposed to a blow-up of the $SU(2)$ singularity. When the dust has settled, though, the two resolution procedures turn out to be completely equivalent. Among the consistency checks performed in~\cite{Braun:2011zm} is a successful computation of the chiral index in models with $SU(2)\times U(1)_X$ and $Sp(2) \times U(1)_X$ gauge symmetry and a match of the D3-brane tadpole with perturbative results in the type IIB limit.

\subsection{Summary of results}

We perform the construction of $U(1)_X$ flux, following our logic spelled out above, for GUT models with gauge group $SU(5) \times U(1)_X$.
The choice of this gauge group is of course motivated by the aim of constructing globally defined F-theory GUT models.
The technical core of the present paper is the resolution of the $SU(2)$ curve responsible for the $U(1)_X$ factor together with the resolution of the $SU(5)$ singularity in a way that gives full access to the resolved fibre over the matter curves. 
In this regard, our work has considerable overlap with the recent analysis in~\cite{Marsano:2011hv} \footnote{Note that~\cite{Marsano:2011hv} focuses on the global realisation of gauge fluxes in the spirit of the so-called spectral divisor construction~\cite{Marsano:2011nn}, which is a different approach to gauge fluxes than the one pursued here.} which independently used a blow-up procedure to resolve an $SU(5)$ model, however, \emph{without} a further $U(1)$ restriction. At a phenomenological level, apart from serving as a welcome selection rule that forbids dimension four proton decay, the $U(1)$ symmetry enhancement leads to a set of GUT singlets with the correct quantum numbers to be interpreted as right-handed neutrinos. The computation of the chirality of such GUT singlets, which are sensitive to the full details of the global compactification, has largely remained elusive in the "semi-local" approach to F-theory GUT model building via spectral covers~\cite{Blumenhagen:2009yv,Marsano:2009wr}. 

The blow-up procedure, performed in this article, reproduces the toric weights of the resolution divisors as appearing in the toric examples of~\cite{Blumenhagen:2009yv,Grimm:2009yu,Chen:2010ts,Knapp:2011wk,Grimm:2010ez}.
The main point of our analysis, however, is to make visible the structure of the fibre above the matter curves as a prerequisite for computing the chiral index. Along the way, we give an explicit procedure to derive the $U(1)_X$ charges of the matter states on purely geometric grounds. These charges identify $U(1)_X$ as the abelian subgroup in the breaking $SO(10) \rightarrow SU(5) \times  U(1)_X$.\footnote{In common abuse of noation, here and in the sequel we say $SO(10)$, but mean "Spin(10)". }
In particular, we identify the ${\bf 10}_{-1}$ and ${\bf \ov 5}_{3}$ as states descending from the spinorial representation of $SO(10)$ while the Higgs ${\bf 5}_{-2} + c.c.$ is of the type present also in perturbative models. This explains why the ${\bf 10 \, 10 \, 5}$ coupling is present in generic F-theory models but not at the perturbative level in Type IIB orientifolds: While the $U(1)$ selection rules operate in exactly the same manner, the charge assignments differ due to the possibility of multi-pronged strings states in F-theory. 

The final result for the chiral index is then extremely simply: The chiral index of a state of $U(1)_X$ charge $q$ that is localised along a matter curve $\cal C$ on the base space is $q \int_{\cal C} F$. In fact, this result matches the formula derived in the context of  the $S[U(4) \times U(1)_X]$ split spectral covers for the special case of zero non-abelian $SU(4)$ bundle part -- however, it matches only for matter states charged under $SU(5)$. These are the fields localised on the GUT brane to which the local philosophy of the spectral cover construction applies. The $SU(5)$ singlets $\mathbf{1}_{5}$ on the curve of $SU(2)$ enhancement away from the GUT brane on the other hand are sensitive to the global details of the compactification. Indeed, the chirality formula derived from the proper $G_4$-flux corrects the spectral cover formula accordingly.
With the explicit resolution at hand it is a simple matter to compute the D3-tadpole $\frac{1}{2} \, \int_{\hat Y_4} G_4 \wedge G_4$ and to evaluate the D-term supersymmetry condition for the gauge flux. Again we find global corrections compared to the spectral cover expressions. This demonstrates that the $G_4$-flux can be viewed as a global extension of the split spectral cover fluxes and that the latter cannot be trusted except for the chirality of the $SU(5)$ matter states, which constitutes a truly local observable.

In addition to the simple form for the chirality in $SU(5) \times U(1)_X$ models, we arrive at slightly more involved expressions for the chiral indices in non-restricted $SU(5)$-models. To this end we partially define $G_4$ in terms of 4-cycles that cannot be represented as the dual of the intersection of two divisors in the four-fold, as was recently described in~\cite{Braun:2011zm}, where, however, the chirality was not computed. Upon extending this we find that the recombination process implicit in moving away from the $U(1)$-locus in complex structure moduli space is nicely reflected in the chirality formula for those curves which are affected by the recombination. In the other cases, the chiral index takes the same form as in the $U(1)$-restricted case.\\

The remainder of this paper is organised as follows:
In section \ref{sec_U(1)restr} we review the details of the $U(1)$-restricted Tate model with special emphasis on singularity resolution via blow-up and the appearance of the $U(1)_X$ gauge potential. 
Section \ref{sec_res} is devoted to a detailed resolution of the $SU(5) \times U(1)_X$ model. 
We begin in \ref{sec:resolution} by discussing the blow-up procedure for the combined resolution of singularities associated with the non-abelian and the abelian part of the gauge group and relate this to the singularity resolution in the framework of toric geometry. In section \ref{p1_structure} we outline the general procedure to deduce the fibre structure in co-dimension one, two and three. The details of this computation are collected in appendix \ref{app_resol}. We proceed in section \ref{sec_MatterCurves} with an in-depth analysis of the fibre structure over the matter curves. In section \ref{sec_U(1)X} we study the geometric realisation of the $U(1)_X$ gauge symmetry and compute the $U(1)_X$ charges of the charged matter fields.
The $G_4$-flux is the subject of section \ref{sec_G4}. 
After stating the Freed-Witten quantisation condition, whose derivation will be presented in the upcoming \cite{appear}, we compute in \ref{sec_chiral}  the chiral index of charged matter states including $SU(5)$ singlets, making heavy use of the geometric structure found in the previous section. Section \ref{sec_global} is devoted to the global consistency conditions such as D3-tadpole and D-term supersymmetry.
In section \ref{sec_compare} we  compare the $G_4$-flux to the split spectral cover construction. In section \ref{sec_model} we illustrate
the use of $G_4$ gauge fluxes in F-theory compactifications by constructing a three-generation $SU(5)\times U(1)_X$ model on a Calabi-Yau four-fold which meets the D3-brane tadpole, the D-term supersymmetry as well as the Freed-Witten quantisation condition.
Finally, in section \ref{sec_a6neq0} we consider the deformation of the $U(1)$-restricted model and its $G_4$-flux by brane recombination and identify the deformed chiral flux.
Many details of the computations of section \ref{sec_res} and \ref{sec_G4} are relegated to the appendices.
Our conclusions are contained in section~\ref{sec_Conc}.

%%%%%%%%%%%%%%%%%%%%%%%%%%%%%%%%%%%%%%%%%%%%%%%%%%%%%%%%%%%%%%%%%%%%%%%%
\section{\texorpdfstring{$U(1)$}{U(1)}-restricted Tate models and their resolution}
\label{sec_U(1)restr}

To set the stage, we present in this section the details of the $U(1)$-restricted Tate model and describe in detail its resolution via blow-ups. This is important in order to understand the resulting  $U(1)$ gauge symmetry.

We consider an F-theory compactification on the elliptically fibred Calabi-Yau four-fold $Y_4$ described by a Weierstrass model in Tate form. The Tate polynomial cuts out $Y_4$ as the hypersurface
\begin{equation}
\label{PT1}
 P_T = \{y^2 + a_1 x y z + a_3 y z^3 = x^3 + a_2 x^2 z^2 + a_4 x z^4 + a_6 z^6\}
\end{equation}\\
of a ${\mathbb P}_{2,3,1}$ bundle over a base space $B_3$.
As usual  $(x,y,z)$ denote the homogeneous coordinates of the ${\mathbb P}_{2,3,1}$ fibre, which can therefore not vanish simultaneously so that $xyz$ lies in the Stanley-Reisner ideal. The $a_i$ depend on the base coordinates in such a way as to form sections of $K_{B_3}^{-i}$, powers of the canonical bundle on $B_3$. Note that the intersection of $P_T$ with the divisor $\{z=0\}$ leads to $y^2 = x^3$. Together with the linear relation from ${\mathbb P}_{2,3,1}$ this fixes a point in the fibre. Thus  $\{z=0\}$ represents a copy of the base $B_3$, the unique section of the generic Weierstrass model.

While we are at it, let us fix some notation:
Unless stated otherwise we denote a divisor defined as the vanishing locus of some coordinate or polynomial $t$ by $\{ t=0\}$ or sometimes short $\{t\}$. Its homology class in $H_6(Y_4)$, or $H_4(B_3)$, will be referred to as $T$, with Poincar\'e dual two-form $[T] \in H^{2}(Y_4)$, or $H^2(B_3)$.
Furthermore, the first Chern class of the tangent bundle of $B_3$ will often be abbreviated as $c_1 := c_1(B_3)$.

The singular fibres of a Weierstrass model $Y_4$ lie over the discriminant locus $\Delta$ in the base $B_3$, famously known to be given by the vanishing of
\bea
\Delta = 4 f^3 + 27 g^2
\eea
with
\bea
&&  f = -\tfrac{1}{48} \left( b_2^2 - 24 b_4 \right), \quad\quad   g = \tfrac{1}{864} \left( b_2^3 - 36 b_2 b_4 + 216 b_6 \right),  \\ \nonumber
&&  b_2 = 4 a_2 + a_1^2,  \quad  b_4 = 2 a_4 + a_1 a_3, \quad b_6 = 4 a_6 + a_3^2.
\eea
As will be exploited in more detail in the next section, specification of the vanishing orders of the sections $a_i$ as prescribed by the Tate algorithm~\cite{Bershadsky:1996nh} along co-dimension one loci in the base $B_3$ engineers non-abelian singularities in the fibre above the respective divisors. Independently of these non-abelian gauge groups along divisors, the presence of a $U(1)$ gauge group not arising as the Cartan of a non-abelian gauge group is associated with a singularity in complex co-dimension \emph{two}.\footnote{There are also other types of geometries realising massless $U(1)$s. Most notably, consider two seven-branes wrapping homologous divisors with vanishing mutual intersection. For seven-branes of the same [p,q]-type,  the same $S^1$ pinches in the elliptic fibre over both seven-branes. Fibreing this $S^1$ between the seven-branes also gives rise to an element in $H^{1,1}(Y_4)$ that is associated with the $U(1)$ gauge potential of the relative $U(1)$, as discussed more recently in~\cite{Braun:2008ua,Braun:2008pz,Grimm:2011tb}. This non-generic case is the four-dimensional analogue of the situation for F-theory on $K3$, where the seven-branes are points on the base of $K3$ and thus automatically homologous. Finally, the appearance of \emph{massive} $U(1)$s as a consequence of certain non-harmonic two-forms was argued for in~\cite{Grimm:2011tb}.}
The simplest type of such models was worked out in~\cite{Grimm:2010ez}  and involves setting the section $a_6 \equiv 0$.
In this case the $U(1)$-restricted Tate hypersurface embedded by (\ref{PT1}) becomes singular 
at $x=y=0=a_3=a_4$. The concrete form of the singularity follows by inspection of the discriminant locus
\bea
  \Delta &=& \Big(a_4  a_3 a_1 \big[ a_1^4 + 8 a_2 \tilde a_1^2  + 16  ( a_2^2 - 6  a_4 )\big]
                       +  a_3^3  a_1  ( a_1^2 + 36  a_2 ) \\ 
&& \phantom{ \big(} -  a_3^2 \big[ a_2  a_1^4 + 8  a_2^2  a_1^2  + 2 (8  a_2^3 + 15  a_4  a_1^2)   
 - 72  a_4  a_2  \big]  \\
  &&\phantom{ \big(} +  a_4^2  \big[ a_1^4 + 8 a_2  a_1^2  + 16  ( a_2^2 - 4 a_4 ) \big] - 27  a_3^4   \Big)\ . 
\eea
From the vanishing degree of order two one infers a curve of $SU(2)$ singularities located at $x=y=0$ in the fibre over the curve
\bea
\label{C34}
C_{34}: \{a_3 = 0\} \cap  \{a_4 = 0\}
\eea
on $B_3$.

The singularity over $C_{34}$ in the four-fold $Y_4$ must be resolved explicitly. 
Let us assume for now that $C_{34}$ is the only singularity of $Y_4$ and thus consider a model without non-abelian gauge groups, reserving the implementation of further non-abelian singularities for the next section.
The discriminant locus is thus a single connected $I_1$-locus with a co-dimension two-singularity along its curve of self-intersection $C_{34}$. The probably simplest possible type of resolution, which is the one applied in~\cite{Grimm:2010ez} and  which we will also explore in this article, is by a standard blow-up procedure.
In this process one introduces a new homogeneous blow-up coordinate $s$ along with the proper transform   $\tilde x, \tilde y$ of the original coordinates $x,y$,
\bea
{y} = \tilde{y} s, \quad\quad {x} = \tilde{x} s.
\eea
Furthermore, one introduces the extra scaling relation 
\bea
(\tilde x, \tilde y, s) \simeq (\lambda^{-1} \tilde x, \lambda^{-1} \tilde y,\lambda  s),
\eea
which follows by requiring that $x$ and $y$ be unchanged under rescaling $s$.\footnote{Note that together with the $\mathbb P_{2,3,1}$ relation for $(x,y,z)$ this can also be brought into the form $(z, \tilde y, s) \simeq (\lambda z, \lambda  \tilde y,\lambda^2 s)$ used in~\cite{Grimm:2010ez}.}

The effect of this blow-up is that, where before the fibre was given by a degree-$6$ polynomial in $\bbP_{2,3,1}$,
it is now given by a degree-$(6,-1)$ polynomial in the following space (relabeling $\tilde x \rightarrow x$ and $\tilde y \rightarrow y$):
\begin{center}
 \begin{tabular}{c | c c c c}
           &  $x$   &  $y$   &  $z$      &  $s$      \\
  \hline
  $Z$      &  $2$   &  $3$   &  $1$      &  $\cdot$  \\
  $S$      &  $-1$  &  $-1$  &  $\cdot$  &  $1$  
 \end{tabular}
\end{center}
The Stanley-Reisner (SR)-ideal of this ``fibre ambient space'' is now generated by $\{xy, zs\}$ and the proper transform of the Tate polynomial becomes
\begin{equation}
 P_T = \{y^2 s + a_1 x y z s + a_3 y z^3 = x^3 s^2 + a_2 x^2 z^2 s + a_4 x z^4\}.
\end{equation}
On the new, resolved Calabi-Yau four-fold $\hat Y_4$ thus created, $\{z=0\}$ still defines a section of the fibre, giving a copy of the base. Furthermore, the divisor $\{s=0\}$ also gives a copy of the base with an additional $\bbP^1$ over the curve
$C_{34} $ defined in \eqref{C34}. This can be seen by considering the restriction of the Tate polynomial to $\{s=0\}$, which is
\begin{equation}
 a_3 y = a_4 x
\end{equation}
after setting $z$ to $1$ as $sz$ is in the SR-ideal. Over all base points away from $C_{34}$ this fixes the fibre entirely; however, on this curve one is left with a $\bbP^1$ parametrised by $[x:y]$.
In other words, above the curve $C_{34}$, $\{s=0\}$ and the Tate polynomial do not intersect transversally, whereas they do over every other base point. Thus the singularity is replaced by a $\mathbb P^1$.

If the original four-fold $Y_4$ is realised as a hypersurface or complete intersection of a toric space, one can arrive at the same scaling relations and proper transform following the toric algorithm of~\cite{Candelas:1996su,Candelas:1997eh}
 as applied more recently to the resolution of Calabi-Yau four-folds in~\cite{Blumenhagen:2009yv,Grimm:2009yu,
Chen:2010ts,Knapp:2011wk}. This method, which is described in more detail at the end of \ref{sec:resolution}, is computationally very powerful and thus particularly well-suited for an efficient treatment of more complicated models, e.g.~in the presence of extra non-abelian singularities. In~\cite{Grimm:2010ez} it was used to resolve $U(1)$-restricted Tate models describing certain $SU(5) \times U(1)$ F-theory GUT models using the geometries of~\cite{Blumenhagen:2009yv,Grimm:2009yu}.

Finally, we stress that our blow-up procedure differs at a technical level from the small resolution performed for $U(1)$-restricted Tate models in the recent work of~\cite{Braun:2011zm}, even though the final results of both approaches match perfectly. \\

The crucial property of the resolved space $\hat Y_4$ is that by construction $h^{1,1}(\hat Y_4)$ has increased by one compared to $h^{1,1}(Y_4)$ due to the new resolution divisor class $S$.
This signals the appearance of a new massless gauge symmetry. Recall that in the language of F/M-theory duality massless brane $U(1)$ symmetries arise from expansion of the M-theory three-form $C_3$ in terms of elements $\nu_i$ of $H^{1,1}(\hat Y_4)$ `with one leg along the fibre and one leg along the base $B_3$'. This means that
\bea
\label{horizontal}
\int_{\hat Y_4}  \nu_i \wedge \pi^*[D_a] \wedge \pi^*[D_b] \wedge \pi^*[D_c] = 0, \quad\quad \int_{\hat Y_4}  \nu_i \wedge [Z] \wedge \pi^*[D_b] \wedge \pi^*[D_c] = 0,
\eea
where $\pi^*[D_i]$ denotes the pull back of the two-form dual to the divisor $D_i\subset B_3$ and $[Z]$ is the two-form dual to the section $\{z=0\}$ with divisor class $Z$.
In fact, a natural candidate for such a two-form is the combination~\cite{Grimm:2010ez}\footnote{This is true for models without non-abelian gauge groups. As we will see later, in presence of extra non-abelian singularities this expression receives modifications (see also~\cite{Braun:2011zm}).}
\bea
\label{twXU1}
\tw_X = - [S] + c_1 + [Z].
\eea

To verify that the first requirement in (\ref{horizontal}) is fulfilled we note that  the topological intersection numbers of $S$ with three base divisors are the same as those of $Z$. 
This is because the additional $\bbP^1$ in $S$ over $C_{34}$ does not occur for \emph{generic} curves in the product of the divisor classes $c_1(K_B^{-3})$ and $c_1(K_B^{-4})$, but only for the vanishing locus of the specific representatives $a_3 = 0 = a_4$.
Furthermore, in a Weierstrass model the section $Z$ is known to satisfy $\int_{\hat Y_4} [Z] \wedge ([Z] + c_1) \wedge \ldots= 0$, which is just right for $\tw_X$ to also satisfy the second requirement in (\ref{horizontal}). 

Note that the construction of $\tw_X$ automatically allows us to write down the $G_4$-flux associated with the $U(1)_X$. Instead of elaborating on this flux in the simple $U(1)$ model, though, we proceed to an in-depth analysis of $SU(5) \times U(1)_X$ models.

%%%%%%%%%%%%%%%%%%%%%%%%%%%%%%%%%%%%%%%%%%%%%%
\section{\texorpdfstring{$SU(5) \times U(1)_X$}{SU(5)xU(1)X} Models, resolution and matter curves}
\label{sec_res}

In this section we extend the $U(1)$-restricted model by an $SU(5)$ singularity in the fibre over a divisor $\{w=0\}$ in the base $B_3$.
The standard procedure to generate an $SU(5)$ singularity  over a divisor $W$ is to fix the vanishing orders of the sections $a_i$ of the Tate polynomial on $\{w=0\}$ according to Tate's algorithm~\cite{Bershadsky:1996nh}.
In addition, we must set $a_6 \equiv 0$, since we are interested in the $U(1)$-restricted version thereof.\footnote{In the non-restricted case, we would impose the vanishing behaviour $a_6= a_{6,5} w^{\, 5}$.}
In summary, the Tate sections are restricted as
\begin{equation}
 a_1 = a_1,\qquad a_2 = a_{2,1}\, w,\qquad a_3 = a_{3,2}\, w^{\,2},\qquad a_4 = a_{4,3}\, w^{\,3}.
\end{equation}
The discriminant now takes the form
\begin{equation}\label{eq:discriminant-expanded}
\Delta = w^{\,5}\,\left(P + Q\,w + R\,w^{\,2} + S\,w^{\,3} + T\,w^{\,4}\right),
\end{equation}
where
\begin{equation*}
\begin{split}
P =&\,\,  a_1^4\,a_{3,2}\,(- a_1\,a_{4,3} + a_{2,1}\,a_{3,2}), \\
Q =&\,\, a_1^2\,(- a_1^2\,a_{4,3}^2 - 8\,a_1\,a_{2,1}\,a_{3,2}\,a_{4,3} - a_1\,a_{3,2}^3 + 8\,a_{2,1}^2\,a_{3,2}^2),\\
R =&\,\, - 8\,a_1^2\,a_{2,1}\,a_{4,3}^2 + 30\,a_1^2\,a_{3,2}^2\,a_{4,3} - 16\,a_1\,a_{2,1}^2\,a_{3,2}\,a_{4,3} - 36\,a_1\,a_{2,1}\,a_{3,2}^3 + 16\,a_{2,1}^3\,a_{3,2}^2, \\
S =&\,\, 96\,a_1\,a_{3,2}\,a_{4,3}^2 - 16\,a_{2,1}^2\,a_{4,3}^2 - 72\,a_{2,1}\,a_{3,2}^2\,a_{4,3} + 27\,a_{3,2}^4, \\
T =&\,\, 64\,a_{4,3}^3.
\end{split}
\end{equation*}
Comparison with~\cite{Bershadsky:1996nh} confirms the presence of the $SU(5)$-singularity over $w$ and furthermore suggests the following enhancement loci:
co-dimension two enhancements occur on the intersection of $\{w=0\}$ with
\begin{itemize}
\item
 $\{a_1\}$, where $\Delta$ vanishes to order $7$, indicating $SO(10)$-enhancement, 
\item  $\{a_{3,2}\}$, and 
\item $\{a_1\,a_{4,3} - a_{2,1}\,a_{3,2}\}$.
\end{itemize}
In the latter two cases $\Delta$ vanishes to order $6$, indicating $SU(6)$-enhancement.
Co-dimension three enhancements occur on 
\begin{itemize}
\item  $\{ a_1\} \, \cap \, \{ a_{2,1}\}$, 
\item  $\{ a_1\} \, \cap \, \{ a_{3,2}\}$, where in both cases $\Delta$ vanishes to order $8$, indicating $SO(12)$- or $E_6$-enhancement, and 
\item $\{ a_{3,2}\} \, \cap \, \{ a_{4,3}\}$, where $\Delta$ vanishes to order $7$, indicating $SU(7)$-enhancement.
\end{itemize}
Note that the splitting of the $SU(6)$-enhancement curve, as well as the appearance of the $SU(7)$-enhancement point, are features of the $U(1)$-restricted model 
that do not occur in non-restricted $SU(5)$-models. Both are of course intimately related to the presence of a $U(1)_X$ gauge symmetry as we will see.

%%%%%%%%%%%%%%%%%%%%%%%%%%%%%%%%%%%%%%%%%%%%%%%%%%%%%%%%%%%%%%%%%%
\subsection{Resolution}\label{sec:resolution}

We next describe in detail the blow-up procedure to resolve the singularities of the above model.
To this aim four exceptional divisors are introduced to take care of the $SU(5)$ singularity in addition to the resolution divisor $S$ from the $U(1)$ restriction. The resolution process can be motivated by the Tate algorithm as is described in section 7 of~\cite{Bershadsky:1996nh}. Restricting ourselves to the $I_n$-branch for the moment and denoting, as before, the divisor defining the GUT surface by $\{w=0\}$, one can summarise the procedure as follows:
Define $x_0 = x$, $y_0 = y$ on the original, singular manifold. Then at each step of the resolution process, a new variable $e_i$ is introduced such that only those monomials with the lowest order in $(x_k, y_l, w)$ remain in $P_T|_{e_i = 0}$ for the current $k$, $l$. If the remaining polynomial factorises into either $x_k\,\tilde{P}$ or $y_l\,\tilde{P}$, one defines new coordinates on the blow-up by $x_{k+1}, y_l$ or $x_{k}, y_{l+1}$ respectively, where $x_{k+1} = x_k / w$, $y_{l+1} = y_l / w$. With the new coordinates the process is then repeated. The algorithm terminates when $P_T|_{e_i=0}$ does not factorise any further.

Then the resolution process turns out as follows, where from each line to the next a relabeling is implicit, losing the $\tilde{}$ over the resolution coordinates in each case,\footnote{The order of the labels of the $e_i$ is chosen such that their intersection structure coincides with the standard root intersection structure (see below).}
\begin{align*}
                            &                  &       (x,  y,  w) &\rightarrow (\tilde{x} e_1,  \tilde{y}   e_1,\tilde{w} e_1), \\
                        y   &\rightarrow y_1 w &\qquad (x,  y_1,w) &\rightarrow (\tilde{x} e_4,  \tilde{y_1} e_4,\tilde{w} e_4), \\
                        x   &\rightarrow x_1 w &\qquad (x_1,y_1,w) &\rightarrow (\tilde{x_1} e_2,\tilde{y_1} e_2,\tilde{w} e_2), \\
   \qquad \qquad \qquad y_1 &\rightarrow y_2 w &\qquad (x_1,y_2,w) &\rightarrow (\tilde{x_1} e_3,\tilde{y_2} e_3,\tilde{w} e_3). \qquad \qquad \qquad 
\end{align*}
This may be summarised as
\begin{equation*}
 (x,y,w) \rightarrow (\tilde{x} e_1 e_4 e_2^{\,2} e_3^{\,2}, \tilde{y} e_1 e_4^{\,2} e_2^{\,2} e_3^{\,3}, \tilde{w} e_1 e_2 e_3 e_4).
\end{equation*}

If we were considering a generic $SU(5)$ model, in which $a_6$ is not set to zero, this would be sufficient.
For $a_6=0$, one has to perform the additional resolution required by the $U(1)$ restriction. As before, this amounts to
\begin{equation*}
 (x,y) \rightarrow (\tilde{x} s, \tilde{y} s).
\end{equation*}
The total resolution process for the $SU(5) \times U(1)$-restricted Tate model can thus be summarised as 
\begin{equation} \label{eqn:resolution_process}
 (x,y,w) \rightarrow (\tilde{x} s e_1 e_4 e_2^{\,2} e_3^{\,2}, \tilde{y} s e_1 e_4^{\,2} e_2^{\,2} e_3^{\,3}, e_0 e_1 e_2 e_3 e_4) .
\end{equation}
Here $\tilde{w}$ was relabeled $e_0$, motivated by the fact that it now denotes the divisor defining the remaining $\pl{}$ fibred over the GUT surface. Whereas, before the blow-up it defined the entire (singular) torus fibration of that surface. 
The proper transform of the Tate equation now becomes
\begin{equation}\label{tate_proper_transform}
 \begin{split}
  y^2\,s\,e_3\,e_4 &+ a_1\,x\,y\,z\,s + a_{3,2}\,y\,z^3\,e_0^2\,e_1\,e_4 \\
                   &= x^3\,s^2\,e_1\,e_2^2\,e_3 + a_{2,1}\,x^2\,z^2\,s\,e_0\,e_1\,e_2 + a_{4,3}\,x\,z^4\,e_0^3\,e_1^2\,e_2\,e_4.
 \end{split}
\end{equation}

Each of the blow-ups induces a new scaling relation by requiring charge invariance of the resolution routine. As discussed already for the simple $U(1)$ model of the previous section, the blow-up $(x, y) \rightarrow (\tilde{x} s, \tilde{y} s)$ induces the divisor class S, and both $x$ and $y$ are not charged under this class as it only appears on the resolved ambient space. Then charge invariance requires $\tilde{x}$ and $\tilde{y}$ to obtain a charge of $-1$ under this new class S, or, put differently, the scaling relation $(\tilde{x}, \tilde{y}, s) \sim (\lambda^{-1} \tilde{x}, \lambda^{-1} \tilde{y}, \lambda s)$ is induced. Combining the relations from all five blow-ups one arrives at a structure for the ambient space of the form displayed in Table \ref{tab:ambient_fibre_space}. Modulo base triangulations this structure allows for 36 triangulations.
\begin{table}
 \begin{center}
  \begin{tabular}{c||c c c c |c c c c|c || c}
             &  $x$      &  $y$      &   $z$       &   $s$       &   $e_1$     &   $e_2$     &   $e_3$     &   $e_4$     &   $e_0$    &  $P_T$   \\
   \hline
   \hline
   $W$       &  $\cdot$  &  $\cdot$  &   $\cdot$   &   $\cdot$   &   $\cdot$   &   $\cdot$   &   $\cdot$   &   $\cdot$   &   $1$      &  $\cdot$ \\
   $\aK$     &  $2$      &  $3$      &   $\cdot$   &   $\cdot$   &   $\cdot$   &   $\cdot$   &   $\cdot$   &   $\cdot$   &   $\cdot$  &  $6$     \\
   \hline
   $Z$       &  $2$      &  $3$      &   $1$       &   $\cdot$   &   $\cdot$   &   $\cdot$   &   $\cdot$   &   $\cdot$   &   $\cdot$  &  $6$     \\
   $S$       &  $-1$     &  $-1$     &   $\cdot$   &   $1$       &   $\cdot$   &   $\cdot$   &   $\cdot$   &   $\cdot$   &   $\cdot$  &  $-1$    \\
   \hline
   $E_1$     &  $-1$     &  $-1$     &   $\cdot$   &   $\cdot$   &   $1$       &   $\cdot$   &   $\cdot$   &   $\cdot$   &   $-1$     &  $-2$    \\
   $E_2$     &  $-2$     &  $-2$     &   $\cdot$   &   $\cdot$   &   $\cdot$   &   $1$       &   $\cdot$   &   $\cdot$   &   $-1$     &  $-4$    \\
   $E_3$     &  $-2$     &  $-3$     &   $\cdot$   &   $\cdot$   &   $\cdot$   &   $\cdot$   &   $1$       &   $\cdot$   &   $-1$     &  $-5$    \\
   $E_4$     &  $-1$     &  $-2$     &   $\cdot$   &   $\cdot$   &   $\cdot$   &   $\cdot$   &   $\cdot$   &   $1$       &   $-1$     &  $-3$    \\
   \hline
   \hline
             &  $-1$     &  $0$      &   $2$       &   $-1$      &   $1$       &   $0$       &   $0$       &   $1$       &   $2$       \\
             &  $0$      &  $-1$     &   $3$       &   $-1$      &   $2$       &   $1$       &   $0$       &   $1$       &   $3$       \\
             &  $\underline{0}$   &   $\underline{0}$   &   $\underline{0}$   &   $\underline{0}$   &   $\underline{v}$   &   $\underline{v}$   &   $\underline{v}$   &   $\underline{v}$   &   $\underline{v}$
  \end{tabular}
 \caption{Divisor classes and coordinates of the ambient space, not including part of the base coordinates and classes. Here $x, y, z, s$ are coordinates of the ``fibre ambient space" of the Calabi-Yau four-fold. Furthermore, within the CY-four-fold, each of the zero loci of the $e_i$ consists of one of the 5 $\pl{}$s fibred over the GUT surface.
For completeness the base classes W and $\aK = [c_1(B)]$ are included.
The bottom of the table is only relevant to torically embedded Calabi-Yau four-folds. It lists a choice for the vectors corresponding to the one-cones of the toric fan. Their relevance is explained below.}
\label{tab:ambient_fibre_space}
 \end{center}
\end{table}

It is also possible to arrive at the summarised resolution process (\ref{eqn:resolution_process}) via different blow-up routes, such as e.g.:
\begin{equation}
 \begin{split}
  &(1):\\ (x, y, e_0) &\rightarrow (x e_1, y e_1, e_0 e_1),\\ (y, e_1) &\rightarrow (y e_4, e_1 e_4),\\ (x, e_4) &\rightarrow (x e_2, e_4 e_2),\\
                (y, e_2) &\rightarrow (y e_3, e_2 e_3),\\ (x, y) &\rightarrow (x s, y s);
 \end{split}
 \qquad
 \begin{split}
  &(2):\\ (x, y) &\rightarrow (x s, y s),\\ (y, s, e_0) &\rightarrow (y e_4, s e_4, e_0 e_4),\\ (s, e_4) &\rightarrow (s e_3, e_4 e_3),\\
               (s, e_3) &\rightarrow (s e_1, e_3 e_1),\\ (s, e_1) &\rightarrow (s e_2, e_1 e_2).
 \end{split}
\end{equation}
The induced set of scaling relations will be different in each case; however, each set is a linear combination of each other set. The above choice, motivated by the Tate algorithm, is used here because in this case only $x$, $y$, $e_i$ and $e_0$ are charged under each of the $E_i$.

While the various possible resolution routes induce equivalent scaling relations, they lead to partially inequivalent triangulations. Each triangulation leads to a different set of coordinates which are not allowed to vanish simultaneously, i.e.\ a different Stanley-Reisner ideal. As is well-known, these constraints can be deduced e.g.~from the requirement that the Fayet-Iliopoulos D-terms of the underlying linear sigma model is positive. Then the differences between the triangulations become clear by considering the Stanley-Reisner ideal. Its generator set  includes the following elements for all triangulations
\begin{equation}\label{srideal_always}
 \{x y,\, x e_0 e_3,\, x e_1 e_3,\, x e_4,\, y e_0 e_3,\, y e_1,\, y e_2,\, z s,\, z e_1 e_4,\, z e_2 e_4,\, z e_3,\, s e_0,\, s e_1,\, s e_4,\, e_0 e_2\}
\end{equation}
along with one of the following 36 options,
\begin{equation}\label{srideal_optional}
 \left\{
  \begin{split}
   & y e_0 \\ & z e_4
  \end{split}
 \right\}
 \otimes
 \left\{
  \begin{split}
   & x e_0,\, x e_1 \\ & x e_0,\, z e_2 \\ & z e_1,\, z e_2
  \end{split}
 \right\}
 \otimes
  \left\{
  \begin{split}
   & s e_2\\ & x e_3
  \end{split}
 \right\}
 \otimes
 \left\{
  \begin{split}
   & e_0 e_3,\, e_1 e_3 \\ & e_0 e_3,\, e_2 e_4 \\ & e_1 e_4,\, e_2 e_4
  \end{split}
 \right\}.
\end{equation}
One notes however that these only lead to six different Calabi-Yau four-folds: From the proper transform of the Tate polynomial (\ref{tate_proper_transform}) and the SR-ideal elements that occur for all triangulations (\ref{srideal_always}), it is clear that all elements of the first two columns of the above list vanish on the Calaby-Yau four-fold. For example, even if $Z$ may intersect $E_1$ in the ambient space, it never intersects any of the $E_i$ on the four-fold. This allows us to fix the elements from those two columns for the future analysis, and we make the canonical choice $z e_4$ and $z e_1,\, z e_2$. Let the remaining 6 triangulations be denoted by $T_{ij}$ with $i \in \{1, 2\}$ and $j \in \{1, 2, 3\}$, so $i$ runs over the third and $j$ over the last column.\\

%%%%%%%%%%%%%%%%%%%%%%%%%%%%%%%%%%%%%%%%%%%%%%
We conclude this discussion with the following comparison with resolutions in the context of toric geometry:
If the Calabi-Yau four-fold is embedded in a toric ambient space, there exists an alternative resolution method, described in~\cite{Candelas:1996su,Candelas:1997eh,Blumenhagen:2009yv,Grimm:2009yu,Knapp:2011wk}. In such a case, the generators of the one-dimensional cones of the toric variety form a rational strictly convex polytope in $d$ real dimensions, where $d$ is the complex dimension of the toric ambient space. Monomials of the Tate polynomial (with the $a_i$ expanded) then correspond to points in the polar dual polytope, the M-lattice polytope (if the Calabi-Yau is a hypersurface) or to points in one element of the nef-partition of the polar dual polytope (if the Calabi-Yau is a complete intersection).\footnote{The latter only holds if such a partition exists.}

Restricting the Tate polynomial coefficients to $a_i = a_{i,k} w^k$ therefore corresponds to removing points from this dual polytope and constructing the dual of the remainder. As was shown in~\cite{Candelas:1997eh}, for the canonical choice of one-cones:
\begin{equation}
 x   = (-1,  0, \underline{0}), \quad\quad y   = ( 0, -1, \underline{0})  \quad\quad z   = ( 2,  3, \underline{0}) \quad\quad  e_0 = ( 2,  3, \underline{v})
\end{equation}
the above algorithm determines the exceptional variables in the $SU(5)$ case to take the form
\begin{equation}
 e_1 = ( 1,  2, \underline{v}), \quad\quad e_2 = ( 0,  0, \underline{v}), \quad\quad e_3 = ( 1,  1, \underline{v}), \quad\quad e_4 = ( 0,  1, \underline{v}).
\end{equation}
Then the scaling relations induced by these one-cones are precisely the ones obtained in the above-mentioned resolution process motivated by the Tate algorithm, and vice versa. This nicely connects the two algorithms and provides a consistency check for the resolution process.

%\subsubsection*{General Remark}
Let us note that the structure of the above resolution ambient space is that of a torically described ambient fibre fibred over a possibly non-toric base three-fold. We have further seen that the blow-up procedure of~\cite{Bershadsky:1996nh}, which holds for general models, produces the same additional scaling relations as  the alogrithm of~\cite{Candelas:1997eh} for torically embedded models. Then for any potential gauge group inducing singularity, one can use Figure 3.2 and Table 3.1 of the toric paper~\cite{Candelas:1997eh} to immediately read off the scaling relations for the fibre ambient space of the resolution, regardless of whether this resolution is toric or not. For the reader's convenience the structure of those scaling relations is summarised in Appendix \ref{app_Ambient}.

%%%%%%%%%%%%%%%%%%%%%%%%%%%%%%
\subsection{Fibre \texorpdfstring{$\pl{}$}{P1} structure}\label{p1_structure}

With the fibre ambient space of the resolution at hand, the natural next step is to investigate the fibre itself. This is well-known to be generically a torus, which splits into 5 $\pl{}$'s over the GUT surface, and is expected to split into more than 5 $\pl{}$'s over enhancement curves and points on said surface. In addition, in the $U(1)$-restricted model the torus also splits into 2 $\pl{}$s over the locus $a_{3,2} = a_{4,3} = 0$. The intersection structure of the GUT-$\pl{}$s is expected to be that of the extended Dynkin diagram $\tilde{A}_4$ associated with $SU(5)$ on the GUT-surface, $\tilde{A}_5$ or $\tilde{D}_5$ on the enhancement curves and $\tilde{A}_6$, $\tilde{D}_6$ or a degenerate version of $\tilde{E}_6$ on the enhancement points. The degeneration of the latter was recently noticed in~\cite{Esole:2011sm} in the framework of a small resolution process, as opposed to blow-up. Indeed, this (with correct $\pl{}$-multiplicities for the $\tilde{D}_i$-cases) is precisely the structure one finds in the framework described thus far (and also in the independent~\cite{Marsano:2011hv}).

Determining the fibre structure is rather technical and we relegate the computations to appendix \ref{app_resol}.
 Here we merely present the general idea and summarise the results of this analysis.\\

First of all the 5 generic $\pl{}$s in the fibre over the GUT surface are given by the fibres of the divisors $\{e_i=0\}, i \in \{0,\ldots,4\}$ inside the Calabi-Yau manifold. Put differently, the $\pl{}$s are given by the intersections
\begin{equation}\label{eqn:general_p1_definition}
 \pl{i} = \{e_i\} \, \cap \, \{P_T|_{e_i = 0}\} \quad \cap \quad \{y_a\} \cap \, \{y_b\}, \qquad i=0,\,\ldots,\, 4
\end{equation}
inside the five-fold which itself is obtained by fibreing the fibre ambient space over the base.\footnote{In the case of toric Calabi-Yau hypersurfaces, e.g.~this is simply the total ambient space; in the case of CICYs (of the form $P_T \cap P_{B_1} \cap ... \cap P_{B_n}$) this is the space given by the intersection of the various $P_{B_i}$ inside the overall ambient space.} Here $\{y_a\}, \{y_b\}$ denote divisors corresponding to base coordinates which are neither $\{e_0\}$ nor any of the enhancement loci, and we assume that their intersection on the GUT-surface is $1$. For general intersection number $n$, (\ref{eqn:general_p1_definition}) defines the formal sum of $n$ $\pl{i}$s.

A standard and well-known property of the resolution ${\mathbb P}^1$s is that they intersect in the fibre according to the extended Dynkin diagram of the gauge group $G$, which in this case is $SU(5)$. More precisely, the intersection pattern of the $\mathbb P^1$-fibred resolution divisors $i=0, \ldots, {\rm rk}(G)$ is
\bea
\int_{\hat Y_4} \pi^*\gamma \wedge [E_i] \wedge [E_j] = - C_{ij} \, \int_{W}  \gamma \quad\quad\quad  \forall  \quad \gamma \in  H^4(B_3) 
\eea
with $C_{ij}$ the Cartan matrix of $G$. We use the sign conventions that $C_{ij}$ has a $+2$ on the diagonal. Indeed these intersection numbers are derived in detail in the appendix.

To see the $\pl{}$-fibre structure above enhancement curves, one notes that $P_T|_{e_i = 0}$ may factorise above certain loci. For example, consider the case $i=1$. Generically, the first ${\mathbb P}^1$  denoted  by $\pl{1}$ is given by
\begin{equation}
 \{e_1\} \cap \, \{y^2\,s\,e_3\,e_4 + a_1\,x\,y\,z\,s\}  \quad \cap \quad \{y_a\} \cap \, \{y_b\}.
\end{equation}
Since $y e_1$, $s e_1$, and $z e_1$ are in the SR-ideal for all triangulations, this can be simplified to
\begin{equation}
 \{e_1\} \cap \, \{e_3\,e_4 + a_1\,x\}   \quad \cap \quad \{y_a\} \cap \, \{y_b\}.
\end{equation}
On $a_1 \neq 0$ one can further use the SR-ideal elements $x e_4$ and $x e_1 e_3$ to express $\pl{1}$ as
\begin{equation}
 \{e_1\} \cap \, \{1 + a_1\,x\}   \quad \cap \quad \{y_a\} \cap \, \{y_b\}.
\end{equation}
On the other hand, on $a_1 = 0$ $\pl{1}$ splits into two $\pl{}$s, namely
\begin{equation}
 \begin{split}
  \pl{13} &= \{e_1\} \cap \, \{e_3\} \quad \cap \quad \{a_1\} \cap \, \{y_a\}, \qquad \textrm{and} \\
  \pl{14} &= \{e_1\} \cap \, \{e_4\} \quad \cap \quad \{a_1\} \cap \, \{y_a\}.
 \end{split}
\end{equation}

Again, this is to be understood as a complete intersection on the five-fold% given by fibreing the ambient fibre over the base
. The key to describing the $\mathbb P^1$s over the curves is to realise that the Tate constraint, which is the second intersection, may factorise over the higher co-dimension loci.

In some triangulations $e_1 e_3$ is in the SR-ideal and no splitting occurs for $\pl{1}$; however, for all triangulations some of the $\pl{i}$ split above the locus $a_1 = 0$, such that a total of 6 different $\pl{}$s appears. Their intersection structure changes to $\tilde{D}_6$, the expected extended Dynkin diagram associated to $SO(10)$. We note that in this framework the correct multiplicities of the $\tilde{D}_6$-diagram appear, which differs from the analysis in~\cite{Esole:2011sm}.\footnote{T.W. thanks Thomas Grimm for pointing this out.} Similarly, above the loci $a_{3,2}=0$ and $a_{2,1} a_{3,2} - a_1 a_{4,3} = 0$ the $\pl{}$s generically split to form $\tilde{A}_5$-structures. The details of the splitting processes for each curve are collected in \ref{app_curves}.

Upon inspection of the enhancement points, one further finds $\tilde{A}_6$-enhancement on $a_{3,2} = a_{4,3} = 0$ and $\tilde{D}_6$-enhancement on $a_{1} = a_{3,2} = 0$ - again with the correct multiplicities.
The appearance of the $\tilde{A}_6$-enhancement is a speciality of the $U(1)$-restricted model and in agreement with the field theoretic expectations, given the localisation of $SU(5)$ singlets along the curve $a_{3,2}=0=a_{4,3}$. 

On the locus $a_1 = a_{2,1} = 0$ on the other hand, 2 pairs of triangulations ($T_{i1}$ and $T_{i3}$) lead to (two different) almost-$E_6$-structures, where one of the multiplicities is not as expected from the $E_6$-Dynkin diagram, while the remaining pair of triangulations ($T_{i2}$) leads to the non-Dynkin type structure, which in~\cite{Esole:2011sm} was named $T_{3,3,3}^{-}$. Again, the details can be found in \ref{app_points}.

\begin{figure} 
 \centering
 \def\svgwidth{\textwidth}
 \executeiffilenewer{splittings.svg}{splittings.pdf}%
 {inkscape -z -D --file=splittings.svg %
  --export-pdf=splittings.pdf --export-latex}%
   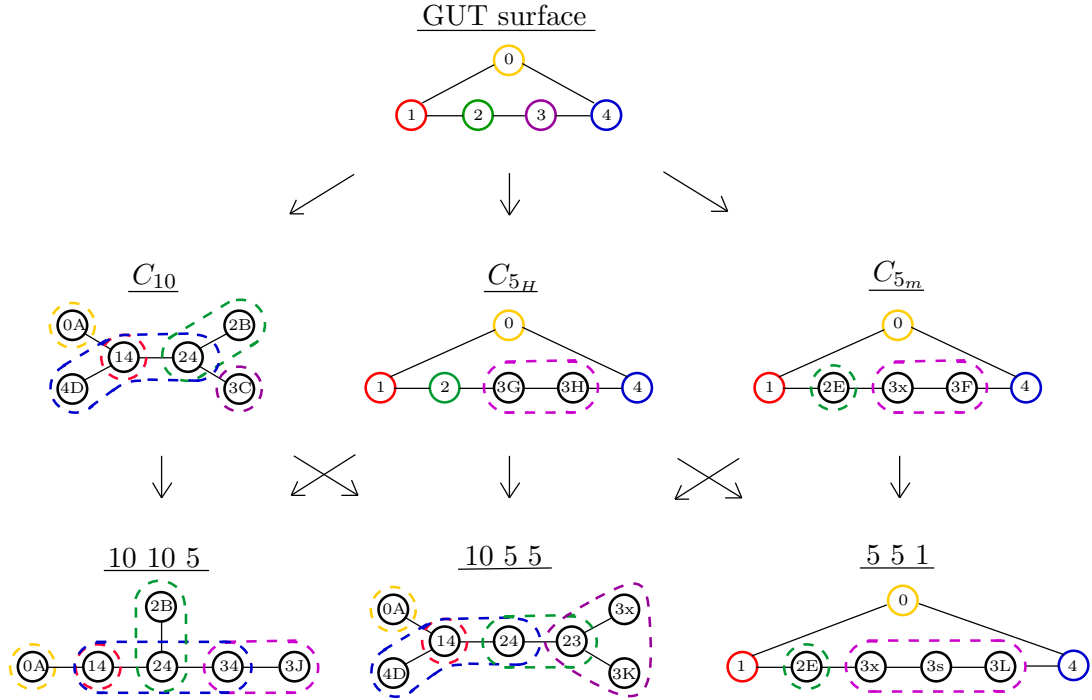%

 \caption{The $\pl{}$-structure and splitting process for triangulation $T_{11}$. The dashed lines encircling one or several $\pl{}$s in the 2nd and 3rd row correspond to the ones of the top diagram, each identified by their colour. Those $\pl{}$s which are marked by a double index with two numbers always have multiplicity $2$, except $\pl{24}$ in the $\mathbf{10}$ $\mathbf{10}$ $\mathbf{5}$-diagram, which has multiplicity $3$. All other $\pl{}$s have multiplicity $1$.}
\label{fig:splittings}
\end{figure}

As an example the $\pl{}$-structure and splitting process for triangulation $T_{11}$ (in the notation introduced after (\ref{srideal_optional})) is depicted in Figure \ref{fig:splittings}.

%%%%%%%%%%%%%%%%%%%%%%%%%%%%%%%%%%%%%%%%%%%%%%%%%%%%%%%%%%%%%%%%%%%%%%%%%%%%%%%%%%
\subsection{\texorpdfstring{$SU(5)$}{SU(5)} matter curves from co-dimension two enhancements}
\label{sec_MatterCurves}

Having understood in detail the ${\mathbb P}^1$-structure of the resolved fibre, we can address the physical interpretation of the co-dimension two loci as matter surfaces~\cite{Katz:1996xe} and of the enhancement points as point of Yukawa interactions~\cite{Donagi:2008ca,Beasley:2008kw}. 
For definiteness the following analysis is carried out for the triangulation $T_{11}$. The remaining cases are covered in appendix  \ref{app_resol}, to which we refer again for most of the technical details.

Let us first recall the general picture  expected to emerge from well-known arguments by F/M-theory duality:
We start with the fields charged under the non-abelian gauge group $G=SU(5)$, beginning in co-dimension one, i.e.\ over the surface $W$ on the base $B_3$.
There are two sources for the gauge bosons in the adjoint representation.
The Cartan generators of the adjoint representation of $G$  are obtained from expansion of the three-form potential into 
the two-forms dual to the 
resolution divisors $E_i, i=1, \ldots, {\rm rk}(G)$.
 In addition, M2-branes wrapping the $\bbP^1$'s of the degenerated fibre can join in all possible ways to form, together with the opposite orientation, the complete set of roots of the Lie algebra. In this picture, the M2-branes wrapping a single $\pl{i}$ are the simple roots $\alpha_i$ of the Lie algebra of $G$.
 
 On co-dimension two loci the singularity enhances further to $\tilde G = SO(10)$ or $SU(6)$. Along these curves some of the $\bbP^1$'s in the fibre split and fuse to form new $\bbP^1$'s with a different intersection pattern. While M2-branes wrapping the `original' $\bbP^1$'s are still present, there are extra massless states from M2-branes  wrapping those new combinations of $\bbP^1$'s. Again, these states include the adjoint representation of $G$. The additional M2-branes wrapping a split $\bbP^1$ can join with the M2-branes wrapping the roots to make up further representations of $G$. 

In co-dimension three there arises yet another enhancement of the singularity in the fibre, and hence, in the resolved manifold additional spheres over these points. Further splittings and fusions occur such that we obtain even more states and, therefore, extra representations at these loci. These points are at the intersection of two enhancement curves. According to the previous argument, the representations before the enhancement always have to be included in the representation at the enhancement. Hence, at these points we have the representations of both curves. The factorisation of  spheres gives us a splitting of states into different representations. Put differently, at these points M2-branes of two possibly different representations can join and form a state in another representation of the group $G$. These gives us the Yukawa couplings to matter localised at the enhancement curves.

After these general remarks, we turn in greater detail to the matter representations in co-dimension two and exemplify how the above picture is realised.
In our notation $C_R$ is the matter surface in the four-fold $\hat Y_4$ associated with representation $R$ of $G$. The projection of $C_R$ to the base $B_3$ is denoted by ${\cal C}_R$. For example, for $R= {\bf 10}$ of $SU(5)$, it turns out  that ${\cal C}_{\bf 10} = \{w=0\} \cap \{a_1=0\}$ on $B_3$. The representation  $R$ is characterised by its highest weight vector $\vec \beta_R^{1}$. Its descendants  $\vec \beta_R^k$, $k=2,\,\ldots,\,\textmd{dim}(R)$, are obtained by acting with the root vectors.  Each of the dim$(R)$ components of the representation $R$ corresponds to a surface $C_R^k$ given by fibreing suitable combinations of $\mathbb P^1$s over  ${\cal C}_R$. The associated physical state is described by a M2-brane wrapping the fibre of $C_R^k$.  From the above we see that $C_R$ splits into various components of this representation.

To see what kind of new representations appear it is most convenient to calculate the Cartan charges of the new states and compare them with the charges as given in weight tables of the representations of $G$, e.g.~\cite{Slansky:1981yr}. Note that this very general procedure has  been applied in various places in the M-theory literature, in particular also in the recent~\cite{Marsano:2011hv}.
Since the gauge bosons in the Cartan of $G$ corresponding to Cartan generator $H_i$ are associated with the two-forms dual to the resolution divisors $E_i$, the Cartan charges of a state are obtained by integrating these two-forms over the two-cycle wrapped by the M2-brane.
Indeed the integrals become the intersection numbers of the curve associated with the state and the $E_i$'s.

Let us therefore intersect the new curves $\pl{\alpha\beta}$ that accrue on the enhancement loci with the divisors $E_i$. To do so, we use the fact that the six-form  $|T^2|$, dual to the generic elliptic curve, is a form entirely on $B_3$,
\begin{align}\label{eq:T2-Ei}
 |T^2|\cdot E_i=0\,,
\end{align}
and that the sum of the split curves has to add up to the class of the original one,
\begin{align}\label{eq:curve-class-conservation}
 |\pl{i}|=\sum_{\alpha,\,\beta} m_{\alpha\beta}\,|\pl{\alpha\beta}|\quad\textmd{and in particular}\quad |T^2|=\sum_{i=0}^4 |\pl{i}|\,.
\end{align}
Here $m_{\alpha\beta}$ is the multiplicity of the split components and $|C|$ is the class to the curve $C$. With \eqref{eq:T2-Ei} and \eqref{eq:curve-class-conservation} we can formulate all products $|\pl{\alpha\beta}|\cdot E_i$ in terms of  effective intersections which we can read off from the equations in appendix~\ref{app_resol}. As an example, let us consider $\pl 4$ in the notation of (\ref{eqn:general_p1_definition}) and its splitting on the \textbf{10}-curve. First of all, from the intersection structure derived in Appendix \ref{app_surface}, we obtain
\begin{equation}
 |\pl 4|\cdot (E_0,\,E_1,\,E_2,\,E_3)=(1,\,0,\,0,\,1)
\end{equation}
for the intersections of $\pl 4$. Together with  \eqref{eq:T2-Ei} and \eqref{eq:curve-class-conservation}, this gives the Cartan charges
\begin{equation}\label{eq:weights-root-4}
 |\pl 4|\cdot (E_1,\,E_2,\,E_3,\,E_4)=(0,\,0,\,1,\,-2),
\end{equation}
which corresponds to the root $\alpha_4$ of $SU(5)$. Similarly, we may obtain the Cartan charges of $\pl 1$,   $\pl 2$, $\pl 3$ which represent, in the obvious way, the other simple roots of $SU(5)$. From appendix~\ref{app_curves}, we see that, for the triangulation we are using here, $\pl 4$ splits into $\pl{14}$, $\pl{24}$ and $\pl{4D}$. Furthermore, from tables \ref{10_1:splitting} and \ref{10_1:root_representation} we observe that $\pl{14}$ has the same Cartan charges as $\pl 1$. To calculate the Cartan charges of $\pl{24}$, we use
\begin{equation}
 |\pl 2|=|\pl{24}|+|\pl{2B}|\,.
\end{equation}
The Cartan charges of $\pl{2B}$ are obtained in the same way as those of $\pl 2$. With~\eqref{eq:curve-class-conservation}, we then not only identify the Cartan charges of $\pl{24}$ but also those of $\pl{4D}$,
\begin{equation}
\left.\begin{array}{c}
 |\pl{24}|\\
 |\pl{4D}|
\end{array}\right\}
\cdot (E_1,\,E_2,\,E_3,\,E_4)
=\left\{\begin{array}{c}
(1,\,-1,\,1,\,-1)\\
(1,\,0,\,0,\,-1)
\end{array}\right.\,.
\end{equation}

By comparing these vectors with the tables of the irreducible representations of $SU(5)$ as listed e.g.~in~\cite{Slansky:1981yr}, we find that M2-branes wrapping
\begin{equation}
\pl{24}=:-\pl{C_\mathbf{10}^{5}}\,,\quad \pl{4D}=:\pl{C_\mathbf{10}^{6}}\quad\textmd{or}\quad \pl{2B} =:\pl{C_\mathbf{10}^{7}}
\end{equation}
are states of the \textbf{10}-representation. 
Indeed, it is possible to identify all of the matter surfaces $C_\mathbf{10}^k$, the result of which is listed in table \ref{10_1:p1_combination}.
Finally, the same type of analysis applied to the two ${\bf 5}$-matter curves in appendix \ref{app_curves} reveals the matter surfaces associated with the fundamental representations, see tables  \ref{5_3:p1_combination} and \ref{5_-2:p1_combination}.

\subsection{\texorpdfstring{$U(1)_X$}{U(1)X} generator, matter charges and \texorpdfstring{$SU(5)$}{SU(5)} singlets}
\label{sec_U(1)X}

Let us now address the extra $U(1)$ gauge group factor which appears in the $U(1)$-restricted model.
We will generalise the construction of the specific two-form $\tw_X$ of eq. (\ref{twXU1}) leading to a $U(1)_X$ gauge potential $A_X$ via $C_3 = A_X \wedge \tw_X$
from the pure $U(1)_X$ model of section~\ref{sec_U(1)restr}.
It is sensible to define $U(1)_X$ to be orthogonal to the Cartan $U(1)$s within $G=SU(5)$.  
To this end let us first recall some elementary group theoretic facts concerning the Cartan $U(1)$s in a non-abelian gauge group.

Since it will turn out that $G=SU(5)$ and the properly defined $U(1)_X$ enjoy an embedding into a higher group $\tilde G$, we phrase this  discussion in the language of some non-abelian group $\tilde G$ containing the original $G$ as a subgroup according to $\tilde G \rightarrow G \times U(1)_X$ for $U(1)_X$ in the Cartan of $\tilde G$. 
To describe $U(1)_X \subset \tilde G$ one  specifies a linear combination 
\bea
X = \sum_{I=1}^{{\rm rk}(\tilde G)} t^I H_I, \qquad {\rm with } \, \,   H_I \, {\rm the \, \, Cartan \, \, generators \, \, of} \, \,  \tilde G.
\eea
Under the decomposition $\tilde G \rightarrow G \times U(1)_X$
the irreducible representations $R'$ of $\tilde G$
decomposes into a direct sum of irreducible representations $R_q$ of $G$ with $U(1)_X$ charge $q$,  $R'=\oplus_q R_q$. 
The weight vector $\vec \beta_{R'}^k$ of the states of an irreducible representation $R'$ can be expanded in terms of the simple roots of $\tilde G$, 
\bea
\vec \beta^k_{R_q}= [\beta^k_{R_q}]^I \alpha_I, \qquad I=1, \ldots, {\rm rk}(\tilde G). 
\eea
Under the decomposition of $R'$, we obtain the $U(1)_X$ charge $q$ of $R_q$ by 
\bea
\label{chargedef}
q = t^I \, C_{IJ} \, [\beta^k_{R_q}]^J =:  t^I \,[\beta_{R_q}^k]_I  \quad\quad\quad \forall k\le \textmd{rk}(R_q)\,.
\eea
Here $C_{IJ}$ is the Cartan matrix of $\tilde G$ and $ [\beta_{R_q}^k]_I  $, with indices downstairs, are the $\tilde G$-Cartan charges of the state corresponding to $\vec\beta_{R_q}^k$.

\begin{figure}
 \centering
 \def\svgwidth{\textwidth}
 \executeiffilenewer{GlobalSetting.svg}{GlobalSetting.pdf}%
 {inkscape -z -D --file=GlobalSetting.svg %
  --export-pdf=GlobalSetting.pdf --export-latex}%
   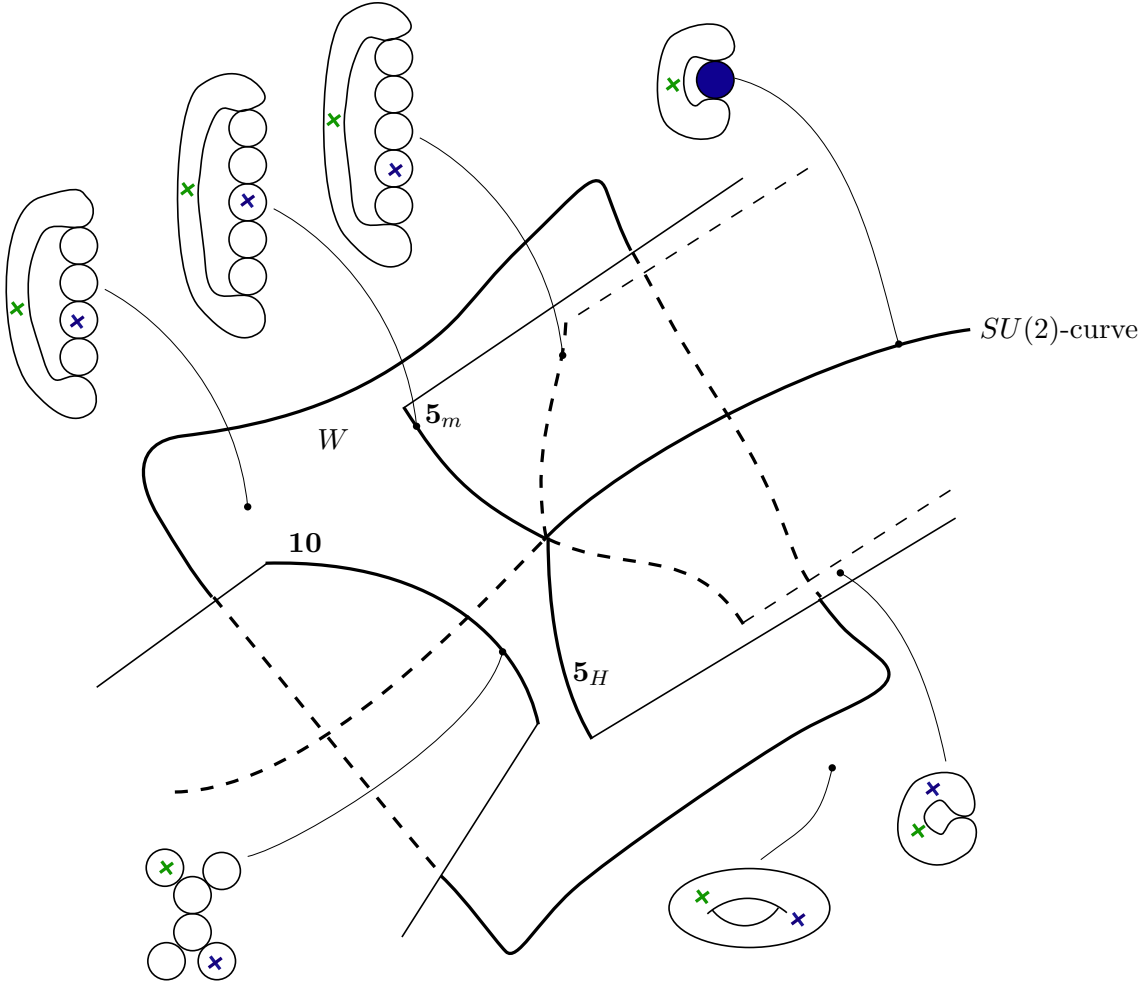%

 \caption{Schematic drawing of the intersection of the generic and degenerate elliptic fibre with the divisors $S$ and $Z$ in the triangulation $T_{11}$. The green and blue crosses indicate the intersection points of $Z$ and $S$, respectively, with the fibre. On the $SU(2)$-curve $S$ itself becomes a $\pl{}$. Note that enhancement points (Yukawa interactions) are ignored in this picture.}
 \label{fig:global-setting}
\end{figure}

To understand how the extra $U(1)$ due to the resolution divisor $S$ fits in, we first
 consider the combination 
 \bea
 \label{E5}
  E_5 :=   S- Z - [c_1(B_3)],
 \eea
which is the na\"ive analogue of the two-form defined in (\ref{twXU1}) that described the $U(1)$ generator in the $U(1)$-restricted model without extra gauge group.
In appendix \ref{app_Intersections} we derive the inter\-section numbers of the resolution divisor $S$ and the remaining divisors of $\hat Y_4$. In Figure \ref{fig:global-setting} we further depict the intersections of $S$ and $Z$ with the various $\pl{}$s.
From these we find that, again in triangulation $T_{11}$ for definiteness,
 \bea
 \label{EiE5intext}
 \int_{\hat Y_4} \pi^*\gamma \wedge [E_i] \wedge [E_5]=|\pl{i}|\cdot E_5\,\int_W \gamma =  \delta_{3,i} \, \int_{W}  \gamma \quad\quad\quad  \forall \quad  \gamma \in  H^4(B_3).
 \eea
Then only $\alpha_3$, the M2-brane wrapping $\pl{3}$, obtains a non-zero charge under the additional $U(1)$.
 More precisely, we can formally 
extend the set of $SU(5)$ Cartan divisors $\{E_i, i=1, \ldots, 4 \}$ by $E_5$ into the set $\{E_I , I=1, \ldots, 5 \}$ and observe the relations, valid for all $\gamma \in  H^4(B_3) $,
\bea\label{eq:rootsSU5cartanSO10}
& &\int_{\hat Y_4}\pi^* \gamma \wedge [E_i] \wedge [E_J] =|\pl{i}|\cdot E_J \,\int_{W}  \gamma=
 \left(\begin{array}{rrrrr}-2&1&0&0&0\\ 1&-2&1&0&0 \\ 0&1&-2&1&\phantom{-}1\\0&0&1&-2&0\end{array}\right) \, \int_{W}  \gamma  \\
&&\textmd{and}\qquad \int_{\hat Y_4} \pi^*\gamma \wedge [E_5] \wedge [E_5] = - 2 \, \int_{B_3}  \gamma \wedge c_1\,.
\eea
The matrix in \eqref{eq:rootsSU5cartanSO10} is, up to the last row and a sign, the \emph{Cartan matrix $C_{IJ}$ of $SO(10)$}.
The fact that the integral involving 2 factors of $E_5$ does not localise on the $SU(5)$  surface $W$ is just as expected, as otherwise we would encounter a fifth root $\alpha_5$ and, therefore, gauge group $SO(10)$ on $W$, not $SU(5)$.   
Nonetheless, the hidden appearance of $SO(10)$ at a group theoretic level is now apparent. 
Indeed, we will see momentarily that the charges of all matter representations do allow for an interpretation in terms of the embedding of $G=SU(5)$ into $\tilde G=SO(10)$. Note that we have \emph{derived} this structure exploiting solely the intersection properties of the resolution divisors computed in detail in the appendix.

From group theory it is known that the $U(1)_X$ generator
\begin{equation}
 X=\sum_{I=1}^5 t^I\, H_I=4\sum_{I=1}^5 C^{I5}\,H_I\qquad\textmd{with}\qquad C^{IJ}C_{JK}=\delta^I_K
\end{equation}
breaks $SO(10)$ to $SU(5)$ by `removing' the simple root $\alpha_5$. All the other roots, $\{\alpha_i,\,i=1,\,\ldots,\,4\}$, are uncharged under this $U(1)_X$.

From this anew group theory interlude, it is obvious that the divisor
\begin{equation}
 \textmd{W}_X=-\sum_{I=1}^5 t^I\,E_I\quad\textmd{with}\quad \vec{t} = (2,4,6,3,5)^T \label{twXSU5}
\end{equation}
is orthogonal to all the (simple) roots of $SU(5)$,
\bea
\label{nocharge}
\int_{\hat Y_4} \tw_X \wedge [E_i] \wedge \pi^*\gamma = |\pl{i}|\cdot \textmd{W}_X\,\int_{B_3} \gamma\wedge [W]=0 \quad\quad \forall \quad \gamma \in H^4(B_3).
\eea
By $\tw_X$  we denote the  dual two-form to the divisor $\textmd{W}_X$.
Up to an  overall factor that determines the normalisation for the $U(1)_X$ charges of the matter states, as will be seen momentarily.

Having $\tW_X$ puts us into a position to compute the $U(1)_X$ charge of the $SU(5)$ matter states in the various representations $R_q$. From the described way how the group theoretic properties are encoded in the geometry, it is clear that the charge $q$ of a state of the representation $R_q$ can be computed by evaluating
\bea
\label{chargeintegral}
|\pl{C_{R_q}^k}|\cdot \tW_X =\sum_{I=1}^5\left(|\pl{C_{R_q}^k}|\cdot E_I\right)\,t^I= q\,. 
\eea 

To calculate this charge we need, besides the already known weights $[\beta_{R_q}^k]_i$, the intersection of $\pl{C_{R_q}^k}$ with $E_5$.
The computation of this intersection proceeds in a manner analogous to the computation of the intersection of the matter $\mathbb P^1$s with the $SU(5)$ divisors $E_i$, detailed in the previous section. It uses  in particular the intersection numbers (\ref{EiE5intext}).

If we add this value as a fifth entry to the Cartan charges of the weight vector we can write the $U(1)_X$ charge as
\begin{equation}
 q_{R}= t^I  [\beta_{R_q}^k]_I.
\end{equation}

For the ${\bf 10}$ representation this procedure leads to
\begin{equation}
 q_{{\bf 10}}= t^I  [\beta_{\bf 10}^k]_I = -1.
\end{equation}
What we should note here is that the extended weights are those of the \textbf{16}-representation of $SO(10)$, to be more specific, those of the $\mathbf{10}_{-1}$ part when $SO(10)$ is broken to $SU(5)\times U(1)_X$. 

Repeating the above steps for the two \textbf{5}-curves $\{a_{3,2} = w=0\}$ and  $\{a_1\,a_{4,3} - a_{2,1}\,a_{3,2} = w=0\}$ curve, we obtain for the $U(1)_X$-charge
\begin{equation}
q_\mathbf{5}=-3\qquad\textmd{and}\qquad q_\mathbf{5}=2\,,
\end{equation}
respectively. Here the $\mathbf{5}_{-3}$ comes again from the \textbf{16}-representation of $SO(10)$ and $\mathbf{5}_{2}$ from its \textbf{10}-representation.

Note that the states descending from the spinorial ${\bf 16}$ representation of $SO(10)$ have no perturbative analogue. In Type IIB language they arise from multi-pronged [p,q]-strings as discussed e.g.~in~\cite{Gaberdiel:1997ud}. The ${\bf 5}_2$ on the other hand does have a perturbative description in terms of fundamental strings.
This of course is in agreement with, and moreover explains, the fact that the ${\bf 10}_{-1} {\bf 10}_{-1} {\bf 5}_2 $ coupling is present in F-theory and not perturbatively in IIB.
While, as pointed out in \cite{Grimm:2011dj,Grimm:2011tb},  the $U(1)$ selection rules operate in exactly the same way as in Type IIB the charge of the states may differ because non-perturbative states are present.

What is still left are the states on the $SU(2)$-curve, $\{a_{3,2}= a_{4,3} = 0\}$. Along this co-dimension two locus the fibre degenerates to two $\pl{}$'s intersecting each other in two points, see appendix~\ref{sec:GenStructureOnC34} for further details. Since the $SU(2)$-curve is normal to the GUT-divisor $W$, the M2-branes wrapping these spheres will not be charged under the Cartans of the $SU(5)$. However, they have a non-zero weight under $E_5$. Using the fact that the generic elliptic fibre is pierced once, both from $S$ and $Z$, we find
\begin{equation}
 (S-Z)\cdot|T^2|=(S-Z)\cdot(|\pl{\mathbf 1_{-5}}| +|\pl{\mathbf 1_{5}}|)=0\,.
\end{equation}
From this relation we can deduce the $E_5$-weight of these states. Their total weight vectors are
\begin{equation}
 |\pl{\mathbf{1}_{-5}}|\cdot E_I=(0,\, 0,\,0,\,0,\,-1)\quad\textmd{and}\quad |\pl{\mathbf{1}_{5}}|\cdot E_I=(0,\, 0,\,0,\,0,\,1)\,.
\end{equation}
This is the missing singlet state plus conjugate of the decomposition of the \textbf{16}-representation of $SO(10)$. Consistently, we find that its $U(1)_X$-charge is 
\begin{equation}
 q_\mathbf{1}=-5\,.
\end{equation}

The spectrum of the $SU(5) \times U(1)_X$ model along with the particle interpretation in the context of an $SU(5)$ GUT model is summarised in table \ref{table-matter}.\footnote{The matter states arising in the $U(1)$-restricted case of~\cite{Braun:2011zm} can similarly be summarised as states in the triplet representation of $SU(3)$. However, the charge normalisation of this paper differs by a factor of~$\tfrac{1}{2}$.}

\begin{equation}\label{table-matter}
 \begin{array}{c|c|c|c}
 \quad\textmd{Matter curve }    {\cal C}_R     \quad &  \qquad R_q   \qquad    & \quad  SO(10) \,  \textmd{origin} \quad  & \textmd{ GUT interpretation}  \\
  \hline
   \{a_1= w=0 \}              &     {\bf 10}_{-1}                                       &   {\bf 16}                     &  (Q_L, U_R^c, e_R^c) \\
     \{a_{3,2} = w=0\}           &{\bf  \ov 5}_3                           &            {\bf 16}               &  (D_R^c, L) \\
   \{a_1\,a_{4,3} - a_{2,1}\,a_{3,2} = w=0\}           & { \bf 5}_2    +   { \bf \ov  5}_{-2}                &   {\bf 10}                                  &  \textmd{Higgs} \\
       \{a_{3,2}= a_{4,3}=0\}                  &   \mathbf{1}_{-5}           &                    {\bf 16}               &   N_R^c
 \end{array}
\end{equation}\\

\section{\texorpdfstring{$G_4$}{G4}-Flux in \texorpdfstring{$U(1)$}{U(1)}-restricted Tate models}
\label{sec_G4}

\subsection{Chiral matter from \texorpdfstring{$G_4$}{G4}-fluxes}
\label{sec_chiral}

After this discussion of resolutions and the structure of the matter curves, we are in the position to approach the construction of a class of chirality inducing gauge fluxes.
What we have achieved towards this aim so far is the construction of a two-form ${\tw_X} \in H^{1,1}(\hat Y_4)$, explicitly
\bea
\tw_X = -t^5 ([S] - [Z] - c_1(B_3)) - \sum_{i=1}^4 t^i [E_i]= - \sum_I^5 t^I [E_I], \qquad t^I = (2,4,6,3,5),
\eea
which  satisfies the constraint (\ref{horizontal}). Upon expansion of the M-theory three-form as $C_3 = A_X \wedge \tw_X$ this realises a $U(1)_X$ gauge potential $A_X$.  
Now, in view of 
\bea
C_3 = A_X \wedge \tw_X + \ldots \quad \Longrightarrow \quad dC_3 = d A_X \wedge \tw_X
\eea
it is clear that this construction of a massless $U(1)_X$ potential automatically yields a natural candidate for the associated gauge flux~\cite{unpublished}. The flux is obtained, as usual, by replacing the three- respectively the four-dimensional field strength $d A_X$ by an internal two-form $F_X \in H^{1,1}(\hat Y_4)$,
\bea
\label{G4}
G_4 = F_X \wedge \tw_X.
\eea
In fact, the well-known condition for an element of $H^4(\hat Y_4)$ to yield a suitable gauge flux is, similar to (\ref{horizontal}), 
\bea
\label{horizontal2}
\int_{\hat Y_4} G_4 \wedge \pi^* [D_a] \wedge \pi^* [D_b] = 0, \quad\quad \int_{\hat Y_4}  G_4  \wedge [Z] \wedge \pi^* [D_a] = 0,
\eea
which is clearly satisfied as long as $F_X \in H^{1,1}(B_3)$. 
In this case also the F-term supersymmetry condition $G_4 \in H^{2,2}(\hat Y_4)$ holds automatically.

The two-form $F_X$ must be quantised in agreement with the M-theory version~\cite{Witten:1996md} of the Type IIB  Freed-Witten quantisation condition~\cite{Freed:1999vc},
\bea
\label{Wittencond}
G_4 + \frac{1}{2} c_2(\hat Y_4) \in H^4(\hat Y_4, \mathbb Z).
\eea
For an analysis of this constraint in the recent F-theory literature see~\cite{Collinucci:2010gz}.
To evaluate this quantisation condition in the case at hand we must compute $c_2(\hat Y_4)$ for the resolved space and analyse its divisability properties ${\rm mod} \,  2$. The details of this computation will be presented in \cite{appear}. Since we will need it for concrete applications, though, we display the final result already here:
\bea
c_2(\hat Y_4) &=& c_2 (Y_4) + \Delta c_2, \\
c_2(Y_4) & =&   c_2(B_3) - c_1^2 + 12 [Z]^2, \\
\Delta c_2 &=& - [W]\,\tw_X + 2\,[W]\,\left\{[Z] + [\aK] - [S] + [E_2] + [E_3]\right\} - 2\,[X]\,[E_3]\\
               && \, + 2\,[\aK] \left\{3\,\left([Z] + [\aK] - [S]\right) - \left(3\,u_i + 2\,v_i - 1_i\right) [E_i] \right\} \nonumber
\eea
 In the above, $c_2(Y_4)$ is the expression referring to the smooth Weierstrass model and $\Delta c_2$ represents the corrections, computed in \cite{appear}, from blow-up of the codimension-one $SU(5)$ and codimension-two $SU(2)$ singularities. The entries of $\vec u= (1,2,2,1)$ and $\vec v = (1,2,3,2)$ are the coefficients of the proper transform of the fibre coordinates $x$ and $y$ with respect to the resolution divisors $E_i$ as displayed in Tabel~\ref{tab:ambient_fibre_space} (and $\vec 1 = (1,1,1,1)$).
One deduces that $\Delta c_2 = [W] \wedge \tw_X \,\, {\rm mod} \, 2$, which, combined with $c_2(Y_4) \in H^2(Y_4, 2 \mathbb Z)$ \cite{Collinucci:2010gz}, yields the quantisation condition
\bea
\label{quantisation-result}
\pi^* F_X + \frac{1}{2} [W] \in H^2(\hat Y_4, \mathbb Z).
\eea

To conclude, the construction of the massless $U(1)_X$ via the $U(1)$-restricted Tate model, as pursued in~\cite{Grimm:2010ez} and in this paper, gives for free a special type of $G_4$ gauge flux. This was independently realised and worked out in great detail also in the recent~\cite{Braun:2011zm}, which has substantial overlap with our work in this regard, even though the logic of the approach and the explicit techniques differ. Note that flux of the type (\ref{G4}) is special in that it can be written as the product of two harmonic forms. We will have more to say about fluxes not sharing this property in section \ref{sec_a6neq0}.

Of course the main motivation to consider gauge fluxes in compactifications  is the fact that they can give rise to a chiral spectrum of charged matter modes.
According to the general expectation (see e.g.~\cite{Donagi:2008ca, Hayashi:2008ba}) the chiral index of an ${\cal N}=1$ chiral multiplet localised on a matter surface should be given by an appropriate integral of $G_4$ over the corresponding locus.
To the best of our knowledge no \emph{derivation} from first principles, i.e.~involving the physics of wrapped M2-branes, of this intuitive assertion has been given. However, it is possible to compare this ansatz with known expressions in dual heterotic or Type IIB setups. Indeed, this is the route we will follow (see also~\cite{Braun:2011zm} and, for a similar treatment of the so-called spectral divisor proposal for gauge fluxes,~\cite{Marsano:2011hv}). 

Consider a matter multiplet in representation $R_q$ localised on the matter surface $C_{R_q}$.
In keeping with the notation of the previous section, we denote by $\vec \beta_R^k$ the weight vector associated with $R_q$ and by $C_R^k$ the corresponding component of the matter surface.
From general arguments~\cite{Donagi:2008ca,Hayashi:2008ba} the chiral index is given by integrating $G_4$ over $C_{R_q}^k$, where each value of $k$ should give the same result.
Since we have the full geometric structure of the matter surfaces $C_R^k$ at our disposal, we can explicitly compute this quantity and check if this is the case.
Since the flux splits as in (\ref{G4}) the integral reduces to
\bea
\label{chiralindex-a}
\chi(R_q)   =   \int_{C_{R_q}^k} \pi^*F_X \wedge \tw_X =      |\pl{C_{R_q}^k}|\cdot \tW_X    \int_{{\cal C}_{R_q}} \imath^*F_X. 
\eea
Now we use the result of  (\ref{chargeintegral}), where we showed that for each value of $k$ the pre-factor just gives the $U(1)_X$ charge $q$ of the state.
To summarise, we have found that
\bea
\label{chiralindex}
\chi(R_q)  = q  \int_{{\cal C}_{R_q}} \imath^*F_X.
\eea
This matches precisely the chiral index derived from the Hirzebruch-Riemann-Roch theorem for matter at the intersection curves of two seven-branes in perturbative Type IIB orientifolds (see e.g.~\cite{Blumenhagen:2008zz} for a discussion and references). As we will discuss in section \ref{sec_compare} it is also consistent with heterotic duality and spectral covers.

Note that the final expression (\ref{chiralindex}) is of course straightforward to evaluate given a concrete Calabi-Yau four-fold $Y_4$ as it only involves an integral of $F_X$ over the curve ${\cal C}_R$ in the base. For the case of the $SU(5) \times U(1)_X$ model the $U(1)_X$ charges and the curves ${\cal C}_R$ of the various matter representations are summarised in (\ref{table-matter}).

\subsection{Global constraints: D3-tadpole, D-term and St\"uckelberg masses}
\label{sec_global}

In this section we describe the global consistency conditions that must be satisfied by the $G_4$-flux. In fact, it is these global aspects for which a full understanding of fluxes in terms of explicit four-forms becomes particularly important.

Turning on $G_4$-flux leads to a contribution in the M2/D3-tadpole condition of the form~\cite{Sethi:1996es}
\bea
\label{NM2}
N_{D3} + \frac{1}{2} \int_{\hat Y_4} G_4 \wedge G_4 = \frac{\chi (\hat Y_4)}{24}
\eea
with $\chi(\hat Y_4)$ the Euler characteristic of the resolved elliptic four-fold. 
Given the concrete expression for $G_4$ in (\ref{G4}) it is a simple matter to compute the flux contribution explicitly,
 \bea
\frac{1}{2} \int_{\hat Y_4} G_4 \wedge G_4 &=& \frac{1}{2} \, \int_{\hat Y_4} \pi^*F_X \wedge \pi^*F_X \wedge \Big(\sum_I^5 t^I [E_I] \Big)  \wedge \Big( \sum_J^5 t^J [E_J] \Big) .
\eea
This is straightforwardly evaluated with the help of the intersection numbers collected in appendix \ref{app_Intersections}, see also \eqref{eq:rootsSU5cartanSO10}.
The result is
\bea
\frac{1}{2} \int_{\hat Y_4} G_4 \wedge G_4 = \int_{B_3} F_X \wedge F_X \wedge   \Big( \frac{Q}{2} \, [W] -{(t^5)}^2 \, c_1(B_3)\Big) 
\eea
with a group theoretic factor
\bea
Q = -\sum_{(I,J)\ne(5,5)}^5   t^J\, C_{IJ} t^J  =  30
\eea
in terms of the $SO(10)$  Cartan matrix $C_{IJ}$. 
The numerical result for this case is then
\bea
\label{G4Tadres}
\frac{1}{2} \int_{\hat Y_4} G_4 \wedge G_4 = \int_{B_3} F_X \wedge F_X \wedge   \left(15 \, [W] - 25 \, c_1(B_3)\right)\,.
\eea

Switching on $U(1)_X$ gauge flux of the form (\ref{G4})  generates a K\"ahler moduli dependent D-term in the four-dimensional ${\cal N}=1$ effective action.
This D-term is rooted in the gauging of the shift symmetry of some of the ${\cal N}=1$ chiral multiplets corresponding to the K\"ahler moduli.
As a result, the $U(1)_X$ symmetry receives a $G_4$-flux dependent St\"uckelberg mass and only remains as a global symmetry in the low-energy effective action.
Both the fact that $U(1)_X$ is not present as a massless gauge symmetry and its persistence as a selection rule broken only by M5-instantons\footnote{Recent investigations of M5/D3-instantons in this context include~\cite{Blumenhagen:2010ja,Cvetic:2010rq,Donagi:2010pd,Grimm:2011dj,Marsano:2011nn,Cvetic:2011gp,Bianchi:2011qh}.} is of course rather important for phenomenological applications. 

The $G_4$-induced gauging, the resulting flux-induced $U(1)_X$ mass and the D-term  can be computed very explicitly via F/M-theory duality.
Their origin in the eleven-dimensional supergravity is the Chern-Simons coupling  $S_{CS} = \frac{1}{12} \int C_3 \wedge G_4 \wedge G_4$.
As shown in detail in~\cite{Grimm:2010ks,Grimm:2011tb}, dimensional reduction of the M-theory action on ${\hat Y_4}$ including $G_4$ flux leads to an action in  3 dimensions which can be brought into the standard form of three-dimensional ${\cal N}=2$ gauged supergravity. Upon uplifting this three-dimensional action to 4 dimensions, one recovers, amongst other things, precisely the form of a D-term potential as well as the flux induced St\"uckelberg masses.

While the reader is referred to~\cite{Grimm:2010ks,Grimm:2011tb} for the details of this dimensional reduction, we here sketch the main ideas for completeness.
One ingredient in the  D-term piece of the three-dimensional scalar potential is the moduli dependent function ${\cal T}$, which in the case at hand  takes the form
\bea
\label{Tfun}
{\cal T} = \frac{1}{4 {\cal V}^2}  \int_{\hat Y_4} J \wedge J \wedge G_4.
\eea
Here $J$ is the K\"ahler form of the Calabi-Yau four-fold $\hat Y_4$ and  $\cal V$ its volume.
The vector multiplet content in 3 dimensions associated with the seven-brane $U(1)$s follows from expansion of 
the K\"ahler form and M-theory three-form as
\bea
J = v^\Lambda \tw_\Lambda + \ldots , \quad\quad C_3 =  A^\Lambda \tw_\Lambda  + \ldots
\eea
with $\tw_\Lambda$ the set of all resolution two-forms introduced by resolution of the various fibre singularities.
The bosonic degrees of freedom of the vector multiplets are $(\xi^\Lambda, A^\Lambda)$ upon rescaling $\xi^\Lambda = \frac{v^\Lambda}{\cal V}$~\cite{Grimm:2010ks}.
The three-dimensional D-term associated with a specific $U(1)_\Lambda$ with potential $A_\Lambda$ is now given by the expression
\bea
D_\Lambda = \partial_{\xi^\Lambda} {\cal T} |_{\xi^\Lambda =0}.
\eea

Applied to the current framework of the $SU(5) \times U(1)_X$ model with $U(1)_X$ flux $G_4$ we are interested in the D-term associated with the multiplet $(\xi^X, A^X)$. From the above we deduce that the D-term of the three-dimensional supergravity is given by $ \frac{1}{2 \cal V} \int_{\hat Y_4} \tw_X \wedge J \wedge G_4$. This form of the D-term for $U(1)_X$ flux was already anticipated in~\cite{Grimm:2010ez}.

To evaluate the appearing integral one performs the same type of computation that leads to the expression for $\frac{1}{2} \int_{\hat Y_4} G_4 \wedge G_4$.
Furthermore one still has to uplift the D-term to the four-dimensional effective action as corresponding to the F-theory limit. This requires rescaling the linear multiplets describing the K\"ahler moduli as detailed in~\cite{Grimm:2010ks}. The effect of this rescaling is to replace the prefactor $\frac{1}{2\, \cal V}$  by   $\frac{1}{ {\cal V}_B}$, where  ${\cal V}_B$ is  the volume of the base space $B_3$.
Taking this into account and combining it with (\ref{G4Tadres}) the result for the D-term of the four-dimensional effective action is
\bea
D_X = -\frac{2}{{\cal V}_B}  \int_{B_3} J \wedge F_X \wedge   \left(15 \, [W] - 25 \, c_1(B_3)\right).
\eea

In the presence of ${\cal N}=1$ chiral matter multiplets $\Phi_i$ charged under $U(1)_X$ the full D-term potential takes the usual form
\bea
V_D \simeq  \Big( \sum_i q_i \,|\phi_i|^2 + D_X \Big)^2.
\eea
If we insist on unbroken gauge symmetry the VEV of the charged matter fields must vanish term by term and the gauge flux must satisfy the D-term supersymmetry condition
\bea
 \int_{B_3} J \wedge F_X \wedge   \left(15 \, [W] - 25 \, c_1(B_3)\right) =0.
\eea
This constraint must be met inside the K\"ahler cone. In absence of any mass terms for the matter fields, however, the D-term $D_X=0$ of course fixes only one linear combination of matter field VEVs and K\"ahler moduli.
Finally note that the $SU(5)$ singlets ${\bf 1}_5 + c.c.$ localised at the matter surface $C_{34}$ of $SU(2)$ enhancement correspond to recombination moduli whose VEV breaks $U(1)_X$ without affecting the $SU(5)$ symmetry.  We will come back to this point in section \ref{sec_a6neq0}.

\subsection{Comparison with split spectral cover bundles}
\label{sec_compare}

It is interesting to compare the globally defined $U(1)_X$ $G_4$-flux with the bundles constructed in the spectral cover approach~\cite{Donagi:2009ra,Hayashi:2008ba}.
The spectral covers encode the geometry of the neighbourhood of the GUT brane $W$. It is therefore expected that they correctly capture local quantities such as the chiral index of $SU(5)$ charged matter, but might miss certain corrections that are sensitive to global details of the four-fold away from the GUT brane. A proposal for a global completion of spectral covers has been made in~\cite{Marsano:2010ix} and, for the case of \emph{non-split} spectral covers, further subjected to global tests in~\cite{Marsano:2011hv}. 
However, since we are interested here in models with $U(1)_X$ symmetry the analogous spectral cover is of the so-called \emph{split} type~\cite{Hayashi:2009ge},~\cite{Marsano:2009gv},\cite{Blumenhagen:2009yv}, generalising the construction of $S[U(N)\times U(1)]$ spectral cover bundles from the heterotic string~\cite{Andreas:2004ja,Blumenhagen:2006ux,Blumenhagen:2006wj}. According to the general arguments of~\cite{Hayashi:2010zp,Grimm:2010ez} such split spectral covers are much more sensitive to the global details of the model and we do not expect to obtain quantitative match at all levels.

To be explicit we compare our fluxes to the $S[U(4) \times U(1)_X]$ bundles in the form described in~\cite{Blumenhagen:2009yv}. 
Such  bundles are constructed in terms of a $U(4)$ bundle $V$ and a line bundle ${\cal L}$ with $c_1(V) + c_1({\cal L}) =0$.
The $U(1)_X$ $G_4$-flux is analogous  only to a special type of such bundles where the non-abelian data of $V$ - associated with the $SU(4)$ piece of the structure group - are switched off.
In the notation of~\cite{Blumenhagen:2009yv}, eq. (85), the corresponding split spectral cover bundle is achieved by setting $\lambda =0$. The flux is then effectively described only by a two-form  on $W$, given in the notation of~\cite{Blumenhagen:2009yv} by the quantity $\frac{1}{4}\, \zeta \in H^{1,1}(W)$ subject to the quantisation condition, eq. (87),
\bea
\label{cond-SCC}
\frac{1}{4} \zeta + \frac12 c_1(W) \in H^2(W, \mathbb Z).
\eea

Our claim is that the $U(1)_X$ $G_4$-flux is the precise global completion of this type of local split spectral cover flux. In particular one must identify $\zeta/4$ with the two-form $F_X$ that appears in $G_4 = F_X \wedge \tw_X$.
From the start it is clear that the $G_4$-flux is more general as $F_X$ need not be an element of $H^2(W)$ but rather of $H^2(B_3)$.
Indeed, inspection of the chirality formulae for the $SU(5)$ charged matter eq. (90), (92), (93) of~\cite{Blumenhagen:2009yv}, confirms that they precisely match our global result (\ref{chiralindex}) if we identify $F_X$ and $\zeta/4$. 
On the other hand, consider the chiral index of  the GUT singlets ${\bf 1}_5$  localised on the curve $C_{34}$ away from the GUT brane. In~\cite{Blumenhagen:2009yv}, eq. (97), a conjecture was made for the chiral index of these singlets (see also~\cite{Marsano:2009wr}), generalising the arguments that had lead to the chiral index of the GUT matter, which in our case reads $\chi({\bf 1}_5)|_{\rm sp. \,  cover} = 5 \int_{B_3} \frac{\zeta}{4} \wedge W \wedge 2 c_1(W)$.  In the globally defined $U(1)_X$ model, the matter curve $C_{34}$ of the singlets is in the class $-12\,c_1(B_3) \wedge c_1(B_3)$. For a general GUT surface $W$, the two expressions do therefore not match. This does not come as a surprise, given the local limitations of the split spectral cover.

Finally consider the D3-brane tadpole, given for the $U(1)_X$ $G_4$-flux of this paper by (\ref{G4Tadres}). The corresponding formula in the split spectral cover, eqn.~(100) of~\cite{Blumenhagen:2009yv}, is $\frac{1}{2} \int_{\hat Y_4} G_4 \wedge G_4  \leftrightarrow - 10 \int_{B_3} \frac{\zeta}{4} \wedge \frac{\zeta}{4} \wedge W$. The reason for the mismatch with the globally correct result (\ref{G4Tadres}) is that the latter receives contributions away from $W$ given by the second term. 
To understand this better note that heuristically we can think of $c_1(B_3)$ as the class of the part of the discriminant locus that meets the GUT brane in the $SO(10)$ curve $a_1 = 0$.\footnote{Recall that $a_1$ is a section of $K_{B_3}^{-1}$.} 
In the spectral cover approach the spectral cover is viewed as a local deformation of the GUT brane obtained by tilting some of its components in the normal direction. This morally describes the component of the $I_1$ discriminant locus in the neighbourhood of the GUT brane.
At a computational level the two "branes" - $\{w=0\}$ and $\{a_1=0\}$ - are not distinguished properly. 
Interestingly, if we indeed identified the classes $W$ and $c_1(B_3)$ in  (\ref{G4Tadres}) we would precisely recover the wrong spectral cover result.\footnote{The same logic almost works - up to an overall factor of 3 - for comparison of the chiral indices for the GUT singlets.}
Similarly we find a mismatch in the quantisation condition (\ref{cond-SCC}) of the spectral cover flux and the result (\ref{quantisation-result}) for the global $G_4$-flux.
Restricted to the GUT surface $W$, the latter gives $ [F_X + \frac{1}{2} (c_1(B_3) + c_1(W))]|_W \in  H^{2}(W, \mathbb Z)$ by adjunction, which again differs by global corrections due to $c_1(B_3)$.

In conclusion we have demonstrated that the $U(1)_X$ gauge flux is perfectly consistent with the interpretation of a proper globalisation of the split spectral cover flux obtained by setting $\lambda=0$. It agrees, where it should, with the spectral cover results and corrects these where the latter are no longer applicable.

\subsection{A three-generation model}
\label{sec_model}

As an illustration we now present an F-theory  $SU(5) \times U(1)_X$ model based on a well-defined Calabi-Yau four-fold and incorporate the $G_4$-flux analysed in this paper to achieve 3 chiral generations of GUT matter.
The example geometry we pick is the model of section 4.1.5 of~\cite{Chen:2010ts}. The base $B_3$ is $\mathbb P^4[3]$ blown-up over two curves and one point. The scaling relations of the homogeneous coordinates on the basis are:
\begin{equation}
 \begin{array}{r|c c c c c c c c}
  & \,\,y_1\,\, & \,\,y_2 \,\,&\,\, y_3 \,\,&\,\, y_4 \,\,&\,\, y_5 \,\,&\,\, y_6 \,\,&\,\, y_7 \,\,&\,\, y_8 \\
  \hline
  H_1\,\, & 1 & 0 & 0 & 1 & 1 & 1 & 1 & 0 \\
  H_2\,\, & 1 & 0 & 0 & 0 & 1 & 0 & 0 & 1 \\
  H_3\,\, & 0 & 1 & 0 & 0 & 0 & 1 & 1 & 0 \\
  H_4\,\, & 0 & 0 & 1 & 0 & 0 & 1 & 0 & 0
 \end{array}
\end{equation}
For later use we note that $[c_1(B_3)] = 2 H_1 + H_2 + 2\,H_3 + H_4$. The intersection form is
\begin{equation}\label{Iform}
 \begin{split}
  I = \;\;& 2\,H_1 H_ 2 H_4+ H_3^3+ H_1H_2^2-2\,H_2^3+ H_1^2 H_3- H_1 H_3^2-2\,H_1^2 H_4\\
        & -2\,H_2 H_4^2 -H_1^3 -2\,H_2^2 H_4+2\,H_1 H_3 H_4-2\,H_3^2 H_4.
 \end{split}
\end{equation}

We take $w \equiv y_1$ as the GUT coordinate and enforce the $SU(5) \times U(1)_X$ restricted Tate model. This choice is motivated by the fact that the brane $\{y_1 =0\}$ is a del Pezzo 4 surface, which makes it suitable for GUT breaking via hypercharge flux as in~\cite{Beasley:2008kw, Donagi:2008kj}.
Since the four-fold is realised as a complete intersection in a toric space, we can perform all computations directly in the framework of toric geometry.\footnote{The unresolved ambient space of the four-fold is given in (146) of~\cite{Chen:2010ts}.}
In particular,  this allows us to compute the Euler characteristic of the resolved $\hat Y_4$ as
\begin{equation}
\label{chivalue}
\frac{\chi ( \hat{Y}_4 )}{24} = \frac{45}{2}.
\end{equation}

In order to construct a well-defined $G_4$-flux we must satisfy the quantisation condition (\ref{quantisation-result}). 
Consistently, this results in a half-integer number for the flux induced three-tadpole $\frac{1}{2} \int_{\hat Y_4} G_4 \wedge G_4$, which, together with (\ref{chivalue}), guarantees an integer number $N_{D3}$ of D3-branes according to (\ref{NM2}).

It is now a simple matter to search for three-generation solutions.
For example, the flux choice
\begin{equation}
 F_X = \frac{1}{2}\left([H_1] + [H_2] + 10\,[H_3] - 8\,[H_4] \right)
\end{equation}
results in the desirable values\footnote{Note that we are counting the chiral index of  ${\bf \ov 5}_m$ as opposed to  ${\bf 5}_m$.}
\begin{equation}
 \chi({\bf 10}) = - \int_{C_{\bf 10}} F_X =  - 3, \qquad \chi({{\bf \ov 5}_m})  = 3  \int_{C_{{\bf \ov 5}_m}} F_X =  - 3, \qquad  \chi({{\bf 5}_H})  = 2 \int_{C_{{\bf 5}_H}} F_X = 0.
\end{equation}
The number of  $SU(5)$ singlets ${\bf 1}_5$ comes out rather large,  $5 \int_{C_{{\bf 1}_5}} F_X = -315$. To see how many of these states can be interpretated as right-handed neutrinos we would have to analyse in detail the Yukawa couplings ${{\bf \ov 5}_m {\bf 5}_H {\bf 1}_5}$. Clearly a detailed phenomenological investigation of this toy model is not what we are aiming for here.
Rather we do note that the D3-brane tadpole induced by the gauge flux is
\begin{equation}
\frac{1}{2} \int_{\hat {Y}_4} \left( \tw_X \wedge F_X \right) \wedge \left(\tw_X \wedge F_X \right)  = \frac{15}{2}.
\end{equation}
The D3-brane tadpole cancellation condition can thus well be satisfied with our flux without the need for anti-D3 branes.
Importantly, the flux induced D-term has a solution $D_X=0$ inside the K\"ahler cone.
Explicitly if we expand the K\"ahler form in terms of  the K\"ahler cone generators  as $J = \sum_i r_i v_1$ with $r_i \in \mathbb R^+$ and
\begin{align}
& v_1 = H_1 + H_3,       && v_2 = H_1 + H_3 + H_4,      && v_3= H_1 + H_2, \nonumber \\
& v_4 = H_1 + H_2 + H_3, && v_5 = 2 H_1 + 2 H_2 - H_4,
\end{align}
the D-term condition
\bea
D_X \simeq 100 r_1 + 190 r_2 + 75 r_3 - 90 r_4  =0 
\eea
 can be solved on a co-dimension one locus inside the K\"ahler cone, as required.

Thus we have found a globally consistent three-generation $SU(5)$ GUT model with supersymmetric flux for which three-brane tadpole cancellation can be achieved by introducing an integer number of D3-branes. More refined model building involving globally defined $G_4$-fluxes along these lines is left for future work.

%%%%%%%%%%%%%%%%%%%%%%%%%%%%%%%%%%%%%%%%%%%%%%%%%%%%%%%%%%%%%%%%%%%%%%%%%%%
\subsection{Generalising the flux to chiral non-restricted \texorpdfstring{$SU(5)$}{SU(5)}-models}
\label{sec_a6neq0}

The $G_4$-flux considered so far is special in the following sense: For a Calabi-Yau four-fold $H^{2,2}(\hat Y_4)$ splits into the so-called vertical and horizontal subspaces $H^{2,2}_{\rm vert}(\hat Y_4)$ and $H^{(2,2)}_{\rm hor}(\hat Y_4)$~\cite{Greene:1993vm}. The difference is that only
an element of $H^{2,2}_{\rm vert}(\hat Y_4)$ can be written as the wedge product of two two-forms.
As noted several times, our $U(1)_X$ flux in the restricted Tate model is precisely of this form.
However, it is related to a chirality-inducing flux not sharing this property upon brane recombination.
To see this we recall that also 
the $U(1)$-restricted Tate model itself is a rather special construction as it involves setting the Tate section $a_6=0$ such as to create a curve of $SU(2)$ singularities along $C_{34}$. 
The massless charged matter states  in representation $\mathbf{1}_5 + c.c.$ localised on  $C_{34}$ are precisely
 the massless recombination moduli which appear as the $U(1)_X$ symmetry is unhiggsed. While this mechanism operates for the most general F-theory models, it has a particularly intuitive interpretation in models with a Type IIB orientifold limit  as discussed in detail in~\cite{Grimm:2010ez}: In this Sen limit the $U(1)$ restriction leads to a split of a single brane brane invariant under the involution (and thus of the type of a Whitney umbrella~\cite{Collinucci:2008pf}) into a brane-image brane pair in the same homology class of the double cover $X_3$ of the F-theory base $B_3$. The $SU(2)$ curve $C_{34}$ is the intersection locus of brane and image-brane not contained in the orientifold plane and indeed hosts the massless recombination moduli charged under the combination $U(1) - U(1)'$ of the brane-image pair. 

Now, Higgsing the $U(1)_X$ symmetry by moving away from the locus $a_6=0$ does not remove the gauge flux completely, nor does it destroy the chirality. Rather, the deformation results in an  $SU(5)$ model with $G_4$-flux that cannot be written as $F_X \wedge \tw_X$ any longer - after all the harmonic two-form $\tw_X$ exists only for $a_6=0$. However, from a field theoretic perspective it is expected that under this deformation the chiral index for the respresentation ${\bf 10}$ along the curve $\{a_1 = 0 = w \}$ is unchanged. The two ${\bf 5}$ curves hosting  ${\bf 5}_3$ and, respectively, ${\bf 5}_{-2}$ in the $U(1)$-restricted case, on the other hand, join into a single object as $a_6 \neq 0$. This corresponds to the fact that as $U(1)_X$ is higgsed there is no distinction between both types of $\bf 5$ matter any more.  The chiral index for matter in representation $\bf 5$ will therefore be the sum of the values for ${\bf 5}_3$ and ${\bf 5}_{-2}$.

It is possible to describe the $G_4$-flux for $a_6 \neq 0$ quite explicitly. In fact, in~\cite{ Braun:2011zm} such type of flux was the starting point
from which the factorisable $G_4$-flux for $a_6=0$ was approached.
Here we treat the problem in the reverse order. As far as the fluxes in the non-restricted model are concerned, our $SU(5)$ setup covers a situation where the non-factorisable flux does induce a chiral spectrum, a case which was not considered explicitly in~\cite{ Braun:2011zm}.

To begin with, we rewrite the flux in the restricted model by introducing the object
 $\tW_R$, with ``R" for remainder,
\begin{equation}
 \tW_R = t^5\,S + \tW_X = t^5(Z + [c_1]) - \sum_{i=1}^4 t^i\, E_i\,.
\end{equation}
For $U(1)$-restricted models we have found $G_4$ to be (partially) given by
\begin{align}
 G_4 &= [(-t^5\,S + \tW_R) \cdot F]              && \textrm{in } \hat{Y}_4    \\
     &= [P_T \cdot (-t^5\,S + \tW_R)  \cdot F]  && \textrm{in } X_5. \label{G4X5a}
\end{align}
Here $X_5$ denotes the ambient five-fold of the resolved Calabi-Yau four-fold in the $U(1)$-restricted case and we rewrite the flux as the class dual to the intersection of the various divisor classes $S, \tW_R, F$.
In non-restricted models there is no class $S$ and in particular there is no second class with the same intersections numbers as $Z$ which could be used to construct the flux in a similar way as above. However, considering the divisor $\{x = 0\}$ one finds that 
\begin{equation}
 P_T|_{x=0} = \left\{ y (y + a_3 z^3) = a_6 \right\}.
\end{equation}
Much in the spirit of~\cite{Braun:2011zm}, this allows us to write a bona fide $G_4$-flux provided $a_6$ factorises as
$a_6 = \rho\,\tau$.  Let us call the four- and five-folds in the case $a_6 \neq 0$ $\tilde Y_4$ and $\tilde X_5$. One obtains two combinations of divisor classes which can be used to define a four-form in $\tilde{Y}_4$ \footnote{Clearly nobody would ever confuse the divisor class ${\cal T}$ with the object appearing in (\ref{Tfun}).},
\begin{align}
 [X \cdot Y &\cdot {\cal P}], \\
 [X \cdot Y &\cdot {\cal T}].
\end{align}
Sticking with ${\cal P}$ for the moment, a consistent type of $G_4$-flux is given by
\begin{equation}
\label{G4new}
 G_4 = [(-t^5\,X \cdot Y + P_T \cdot \tW_R) \cdot {\cal P}] \quad \textrm{in} \quad \tilde X_5.
\end{equation}
Comparison with (\ref{G4X5a}) shows that the general form of the flux for $a_6 \neq 0$ is related to the form of the flux in the $U(1)$-restricted case  by replacing
\begin{align}
\label{ids}
 F \leftrightarrow {\cal P},               \quad \quad\quad 
 P_T \cdot S \leftrightarrow X \cdot Y.
\end{align}
We stress that unlike for $a_6=0$, the $G_4$-flux (\ref{G4new}) factorises only on the ambient $\tilde X_5$ but not on the four-fold $\tilde Y_4$.

The first of the formal identifications in (\ref{ids}) implies that, while in the $U(1)$ restricted model one is free to choose any $F \in H^{1,1}(B_3)$, if $a_6 \neq 0$ only 2-forms $ {\cal P}$ with the property  $0 < {\cal P} < A_6$ can appear.
The second identification guarantees that the expression in terms of $\cal P$ for the chiral index of all states except for the recombination modes along $C_{34}$ is the same as the one in terms of $F$ if $a_6=0$. Indeed, since $s=0$ is the $\pl{}$ pasted into the point $x = y = 0$ in the fibre ambient space (over every base point)\footnote{Upon intersection with $P_T$, the $\mathbb P^1$ remains in $\hat{Y}_4$ only over the curve $C_{34}$.}, the intersections of $S$ with the various $\pl{}$s after the $U(1)$-resolution are the same as those of $X \cap Y$ before the resolution - provided $X \cap Y$ lies on the Calabi-Yau. The latter is the case on the locus $\{\rho =0\}$, which explains the further restriction of $F = {\cal P}$. More concretely one has
\begin{equation}
 \int_{X_5} [P_T \cdot S \cdot {\cal P} \cdot D_a \cdot D_b] = \int_{\tilde{X}_5} [X \cdot Y \cdot {\cal P} \cdot D_a \cdot D_b].
\end{equation}
where $D_a, D_b \neq S$ as $S$ does not exist on $\tilde{X}_5$.

In particular consider the $\pl{}$s in the fibre over the various enhancement curves $\cal{C}_R$, where $R = \mathbf{10}, \mathbf{5}_H, \mathbf{5}_m$. These are given by intersections of the type
\begin{equation}
 e_i \cap P_T|_{e_i=0} \cap {\cal D}_R
\end{equation}
where e.g.~${\cal D}_{{\bf 10}} = a_1$. Then the integral of the flux part $X \cdot Y \cdot {\cal P}$ over a $\pl{}$ fibred over $\cal{C}_R$ is given by
\begin{equation}
 [e_i \cap P_T|_{e_i=0} \cap {\cal D}_R]  \cdot X \cdot Y \cdot {\cal P},
\end{equation}
which is the intersection of $6$ divisors on a five-fold. Note that on generic points of the GUT surface $P_T|_{e_i=0}$ becomes redundant as it does not intersect $x \cap y \cap \rho$ transversally. However, over the enhancement curves, $P_T|_{e_i=0}$  splits in some cases - as analysed in detail in previous sections - and therefore does not always become redundant. Then the flux contribution from $[X \cdot Y \cdot {\cal P}]$ may be nonzero only for those $\pl{}$s for which the second defining equation does indeed become redundant.

To see this more explicitly, consider the $\pl{}$-fibre structure of the non-restricted model. As is summarised in Appendix \ref{app_non-restricted}, this is very similar to the structure of the restricted model: For each of the co-dimension-one and co-dimension-two singular loci, the defining equations for the fibre-$\pl{}$s only change for one of the $\pl{}$s, whilst the others are defined in the same manner. The $\pl{}$s which differ turn out to be precisely those, which are intersected by the divisor $s=0$ in the $U(1)$-restricted case. 
Then from the structure of the other $\pl{}$s it is clear that over generic points along the $SU(5)$ locus only $\pl{3}$ is intersected by $x \cap y\cap \rho$, while over the 
 $\mathbf{10}$-curve only $\pl{3C}$ is intersected (for all other $\pl{}s$, $x$, $y$ or both are contained in the list of variables that have to be non-zero). We thus obtain that for each weight component ${C}^k_{10}$ of the $\mathbf{10}$ matter surface
\begin{equation}
\chi({\bf 10}) =  \int_{C^k_{10}} G_4 = -1 \int_{{\cal C}_{10}} [{\cal P}].
\end{equation}

For the analysis of the recombined $\mathbf{5}$-curve consider the intersection of $\rho$ with the $\mathbf{5}$-curve. Even though the $\mathbf{5}$-curve is a single connected object in the generic non-restricted model, this intersection splits into the two (co-dimension three) loci $\rho \cap a_{3,2}$ and $\rho \cap a_{2,1} a_{3,2} - a_1 a_{4,3}$. One notes that $\pl{3}$ splits differently above the two loci, as is shown in Appendix \ref{app_non-restricted}. In both cases $x \cap y \cap \rho$ intersects only one $\pl{}$, namely the one which is structurally the same as the one intersected by $s=0$ for restricted models. Then the integral of $G_4$ over the matter surface $C^k_5$, for each $k$ corresponding to one of the weights, is the sum of the integrals over the two loci, and the resulting chirality formula becomes
\begin{equation}
 \begin{split}
 \chi({\bf 5}) = \int_{C^k_{\mathbf{5}}} G_4 \quad &= \quad \int_{C^k_{\mathbf{5}_A}} G_4 + \int_{C^k_{\mathbf{5}_B}} G_4    \\[5pt]
                                    \quad &= \quad 2 \int_{{\cal C}_{\mathbf{5}_A}} [{\cal P}] \, - \, 3 \int_{{\cal C}_{\mathbf{5}_B}} [{\cal P}],
 \end{split}
\end{equation}
where ${\cal C}_{\mathbf{5}_A} = \{a_{3,2} =0\} \cap \{w=0\}$ and ${\cal C}_{\mathbf{5}_B} = \{a_{2,1} a_{3,2} - a_1 a_{4,3} =0\}  \cap \{w=0\} $. This is the geometric incarnation of our previous field theoretic statement that the $\ov{\mathbf{5}}_{3}$ and the $\mathbf{5}_{-2}$ states of the restricted model pair up once $U(1)_X$ is broken.\\

Let us conclude this discussion with some general remarks. As stressed before, for $a_6 \neq 0$ the $G_4$-flux is an element of $H^{2,2}(\tilde Y_4)_{\rm hor.}$. For flux of this type, the D-term potential, which was $ \frac{1}{2 \cal V} \int_{\hat Y_4} \tw_X \wedge J \wedge G_4$ in the $a_6 =0$-case, now  automatically vanishes identically and does not put any restrictions any longer on the K\"ahler moduli. This is of course in agreement with the Higgsing of the $U(1)_X$. On the other hand, $G_4$-flux in $H^{2,2}(\tilde Y_4)_{\rm hor.}$ does induce a superpotential of Gukov-Vafa-Witten type $\int_{\tilde Y_4} \Omega \wedge G_4$, whose associated F-term supersymmetry condition famously fixes some of the complex structure moduli of the Calabi-Yau four-fold. This is reflected in the necessity to factorise $a_6 = \rho \tau$~\cite{Braun:2011zm}  in order for the $G_4$-flux (\ref{G4new}) to exist.
It is interesting to compare the situation with the description of gauge fluxes in the weak coupling limit corresponding to Type IIB orientifolds on a Calabi-Yau three-fold $X_3$. 
As always the $I_1$-locus of the F-theory model splits into the orientifold plane together with a single seven-brane. In models without non-abelian gauge enhancement, for $a_6 =0$ this latter seven-brane has the topology of a brane-image brane pair~\cite{Grimm:2010ez}, while for $a_6 \neq 0$ it corresponds to a single invariant brane of Whitney type~\cite{Collinucci:2008pf}. This is expected to hold also in more complicated setups.
In Type IIB, gauge flux $F$ can in general be decomposed into a sum of two types of fluxes $f_1 + f_2$ where $f_1 \in {\iota}^*H^{1,1}(X_3)$, i.e.\ it is the pullback of a 2-form on $X_3$, and $f_2$ is dual to 2-cycle of the brane which is homologically trivial on $X_3$. 
Only fluxes of type $f_1$ can induce a chiral spectrum, while ony fluxes of type $f_2$ can induce a superpotential.
Note that for invariant branes, the allowed gauge flux must be anti-invariant under the orientifold involution, $F \in H^{1,1}_-(D)$. In this case the D-term vanishes identically as observed also in F-theory. It is therefore natural to suspect that chirality inducing $G_4$-fluxes of the type (\ref{G4new}) for $a_6 = \rho \tau \neq 0$ corresponds, in an orientifold limit, to gauge flux $F$ with non-trivial components both of the type $f_1$ and $f_2$. It will be interesting to investigate this more quantitatively in the future. 
Finally we stress that while conceptually very rewarding the generic $SU(5)$-model without any $U(1)$ restriction is phenomenologically less relevant because of  dangerous dimension-four  proton decay operators.

\section{Summary and Outlook}
\label{sec_Conc}

In this article, we have constructed globally defined gauge fluxes in F-theory compactifications based on the $U(1)$-restricted Tate model of~\cite{Grimm:2010ez}.
While the construction of such fluxes is more general, we have focused on the special case of an $SU(5) \times U(1)_X$ model. These are the models that have recently appeared in the context of phenomenology inspired F-theory GUT compactifications.
A special restriction of the complex structure moduli of the elliptically fibred Calabi-Yau four-fold induces a curve of $SU(2)$ singularities in addition to the $SU(5)$ singularities along the GUT divisor. We have detailed the resolution of this singularity via a blow-up procedure resulting in an extra harmonic two-form.
This two-form is used to constructed the form $\tw_X$.
The  $G_4$-flux is then simply $G_4 = \pi^* F_X \wedge \tw_X$ with $F_X \in H^{1,1}(B_3)$.  Such fluxes had been proposed already in~\cite{unpublished} and were also considered independently in the recent~\cite{Braun:2011zm}. 
The technical core of the present work is an analysis of the resolved fibres over the matter curves and the Yukawa points, taking into account both the $SU(5)$ singularity resolution and its interplay with the resolution divisor associated with the $U(1)_X$ gauge factor. Our results precisely match the toric resolution techniques applied to elliptic four-folds in~\cite{Blumenhagen:2009yv,Grimm:2009yu, Chen:2010ts,Knapp:2011wk}. In particular, we derive from first principles the $U(1)_X$ charges of the various matter fields and identify these charges as consistent with the branching $SO(10) \rightarrow SU(5) \times U(1)_X$. Part of the matter arises from the spinor and the remaining part from the vector representation of $SO(10)$. Along the way we confirm the general structure of enhancements which had been found --- albeit with different methods from ours and for the case of the non-restricted SU(5) model without extra $U(1)_X$ symmetry --- recently in~\cite{Esole:2011sm} and~\cite{Marsano:2011hv}.
 
As a consequence of the geometric structure of the matter surfaces, we derive the simple and intuitive formula $\chi(R_q) = q\,\int_{{\cal C}_{R_q}} F_X$ for the chiral index of states of $U(1)_X$ charge $q$ along the curve ${\cal C}_{R_q}$ on the base. This formula holds, in particular, for the chiral index of $SU(5)$ singlets. As these are not localised on the GUT brane, they are especially sensitive to the global details of the four-fold. 
Further, important quantities that can only be determined reliably in a global context are the flux induced D3-brane tadpole, the St\"uckelberg mass for the $U(1)_X$ gauge boson and the D-term supersymmetry condition on the $G_4$-flux, each of which we discuss in detail.

The technology of the $U(1)$-restricted Tate model and $G_4$-fluxes of the type discussed here and, independently, in~\cite{Braun:2011zm} paves the way for truly global F-theory model building free of any assumptions concerning the validity of a spectral cover approach. In fact, we have shown that the $U(1)_X$ fluxes  of this paper represent the global extension of a certain type of split spectral cover fluxes. The expressions for the chiral index for $SU(5)$ GUT matter agree in both cases, while those for the index of the GUT singlets, the D3-brane tadpole and the D-term receive non-local corrections.
Indeed, given an explicit Calabi-Yau four-fold, e.g.~of the type as the models in~\cite{Blumenhagen:2009yv,Grimm:2009yu,Chen:2010ts,Knapp:2011wk}, it is a simple matter to search for consistent $G_4$-fluxes that lead to a chiral spectrum.
We have exemplified this by providing a fully consistent supersymmetric, tadpole canceling three-generation $SU(5)$ GUT model based on a geometry  of~\cite{Knapp:2011wk}.

In providing this model  we have anticipated the explicit form, derived in detail in the upcoming \cite{appear}, of the flux quantisation condition. In this work we will also derive an analytic expression for the Euler characteristic of the $U(1)_X$ restricted model.

Furthermore, we have given a form of the gauge flux for more general, non-restricted models in the spirit of~\cite{Braun:2011zm}, which directly relates to the flux in the $U(1)$-restricted case. The resulting flux remains to be chiral but cannot be understood as a wedge of two two-forms. It will be interesting to analyse in more detail the relation to Type IIB fluxes in the future.

\subsection*{Acknowledgements}

We are grateful to Andreas Braun, Andr\'{e}s Collinucci, Thomas Grimm, Arthur Hebecker, Max Kerstan, Eran Palti and Roberto Valandro for discussions.  T.W. thanks Thomas Grimm for initial collaboration and for informing us about a related project, and acknowledges hospitality of the Max-Planck-Institut f\"ur Physik, M\"unchen. S.K.\ thanks the Klaus-Tschira-Stiftung for financial support. This work was furthermore supported by the Transregio TR33 "The Dark Universe".

\newpage
%%%%%%%%%%%%%%%%%%%%%%%%%%%%%%%%%%%%%%%%%%%%%%%%%%%%%%%%%%%%
%%%%%%%%%%%%%%%%%%%%%%%%  Appendix  %%%%%%%%%%%%%%%%%%%%%%%%
%%%%%%%%%%%%%%%%%%%%%%%%%%%%%%%%%%%%%%%%%%%%%%%%%%%%%%%%%%%%
\appendix
\section{The \texorpdfstring{$\bbP^1$}{P1}-fibre structure of \texorpdfstring{$U(1)$}{U(1)}-restricted \texorpdfstring{$SU(5)$}{SU(5)}-models}
\label{app_resol}
In this appendix we collect several aspects of the $\pl{}$-fibre structure of $U(1)$-restricted $SU(5)$-models such as
\begin{itemize}
 \item the defining equations (partially in inhomogeneous form) of the various $\pl{}$s,
 \item the splitting structure,
 \item the intersection structure of the $\pl{}$s,
 \item the root assignment for each $\pl{}$,
 \item the $\pl{}$-combination for
 \begin{itemize}
  \item each state of the $\mathbf{10}$- or the $\mathbf{5}$-representation on each enhancement curve,
  \item the roots $\alpha_i$ and the highest weights $\mu_{5}$, $\mu_{10}$ on the enhancement points.
 \end{itemize}
\end{itemize}
The general idea for how to find the $\pl{}$s is spelled out in section \ref{p1_structure}. To keep this appendix self-contained, however, we reproduce parts of this description here.\\

Away from enhancement loci, the $\pl{}$s in the fibre over the GUT surface are given by the intersection of the following four divisors inside the five-fold consisting of the fibre ambient space fibred over the base,
\begin{equation}
 \pl{i} = \{e_i\} \, \cap \, \{P_T|_{e_i = 0}\} \quad \cap \quad \{y_a\} \cap \, \{y_b\}, \qquad i=0,\,\ldots,\, 4.
\end{equation}
Here $\{y_a\}, \{y_b\}$ denote base divisors which are neither $\{e_0\}$ nor any of the enhancement loci, and we assume that their intersection on the GUT-surface is $1$. (For general intersection number $n$, the above defines the formal sum of $n$ $\pl{i}$s.) To keep this appendix as clear as possible, in the following we leave out the ``divisor brackets'', $\{\}$, and lose the $y_k$, which are to be thought of as being implicitly present. Then the generic 5 $\pl{}$s are given in homogeneous form by
\begin{align*}
\pl{0}: \quad & e_0 \, \cap \, y^2\,s\,e_3\,e_4 + a_1\,x\,y\,z\,s - x^3\,s^2\,e_1\,e_2^2\,e_3, \\
\pl{1}: \quad & e_1 \, \cap \, y^2\,s\,e_3\,e_4 + a_1\,x\,y\,z\,s, \\
\pl{2}: \quad & e_2 \, \cap \, y^2\,s\,e_3\,e_4 + a_1\,x\,y\,z\,s + a_{3,2}\,y\,z^3\,e_0^2\,e_1\,e_4, \\
\pl{3}: \quad & e_3 \, \cap \, a_1\,x\,y\,z\,s + a_{3,2}\,y\,z^3\,e_0^2\,e_1\,e_4 - a_{2,1}\,x^2\,z^2\,s\,e_0\,e_1\,e_2 - a_{4,3}\,x\,z^4\,e_0^3\,e_1^2\,e_2\,e_4, \\
\pl{4}: \quad & e_4 \, \cap \, a_1\,x\,y\,z\,s - x^3\,s^2\,e_1\,e_2^2\,e_3 - a_{2,1}\,x^2\,z^2\,s\,e_0\,e_1\,e_2. \\
\end{align*}

Let us proceed by considering $\pl{1}$ to illustrate some concepts. Since $y e_1$, $s e_1$ and $z e_1$ are in the SR-ideal for all triangulations, the defining equations can be simplified to
\begin{equation*}
 e_1 \cap \, e_3\,e_4 + a_1\,x.
\end{equation*}
On $a_1 \neq 0$ it is convenient to further use the SR-ideal elements $x e_4$ and $x e_1 e_3$ to re-express $\pl{1}$ and to collect the variables which were set to $1$,
\begin{align*}
 &\qquad \qquad \qquad \qquad &&e_1 \, \cap \, 1 + a_1\,x,   & (y,z, s, e_3, e_4) &= \underline{1}.
\end{align*}
This is what we call the partially inhomogeneous form.
From this one can immediately read off that $\pl{1}$ cannot intersect $\pl{3}$ or $\pl{4}$ as those require $e_3$ or (respectively) $e_4$ to vanish. This reduces the number of possible intersections one needs to consider when determining the intersection structure. The other major property important for the latter aspect is the fact that $\pl{i} \cap \pl{j}$ should be given by the intersection of at most $5$ divisors inside the five-fold, two of which are $\{y_a\}$ and $\{y_b\}$. Then in the above notation, if the four polynomials one obtains considering two $\pl{}$s do not contain a redundant element, then the two do not intersect.\\

Staying in the above example, on $a_1 = 0$ $P_T|_{e_1=0}$ factorises and therefore $\pl{1}$ splits into
\begin{align*}
  &\pl{13}: && e_1 \cap \, e_3,                                                           & (x, y, z, s) &= \underline{1},       \\
  &\pl{14}: && e_1 \cap \, e_4 ,                                                          & (x, y, z, s) &= \underline{1}.
\end{align*}
In some triangulations $e_1 e_3$ is in the SR-ideal and no splitting occurs for $\pl{1}$; however, for all triangulations some of the $\pl{i}$ split above the locus $a_1 = 0$, such that a total of 6 $\pl{}$s appears. In the later subsections of this appendix, for each enhancement locus, we list all $\pl{}$s occurring in some triangulation, summarise the splitting process for each triangulation and determine the intersection structure in each case.\\

With this information at hand one can calculate the Cartan charge of each of the $\pl{}$s. The Cartan charges are minus the intersection numbers with the divisors $E_i$ obtained by fibreing $\mathbb P^1_i$ over the GUT divisor $W$. Again, to stay in the above example, for triangulations in which $\pl{13}$ exists, one has $\pl{1} \rightarrow \pl{13} + \pl{14}$. Then the first Cartan charge for each $\pl{}$ is given by the sum of the intersections with $\pl{13}$ and $\pl{14}$. On the other hand, for triangulations where $\pl{1} \rightarrow \pl{14}$, it is given by the intersection with $\pl{14}$ only. The Cartan charge then allows one to find an expression for each of the $\pl{}$s as a linear combination of the roots, $\alpha_i$, and the highest weights of the $\mathbf{5}$- and the $\mathbf{10}$-representations, $\mu_5$, $\mu_{10}$ respectively. To see how the Cartan charges and the weights are related we refer e.g.~to~\cite{Slansky:1981yr}.

In turn this can be used to determine which $\pl{}$s an M2-brane must wrap in order to lead to a state in a certain representation. We therefore list, for each enhancement curve, the root combination for each $\pl{}$ and the $\pl{}$-combination for each state of the appropriate representation. In the case of the enhancement points, states of both representations occur and for compactness we only list the $\pl{}$-combinations for the roots, $\alpha_i$, and the highest weights, $\mu_5$, $\mu_{10}$.

\subsection{GUT Surface}
\label{app_surface}
The $\pl{}$s, in partially inhomogeneous form, are given by
\begin{align}\label{surface:p1s}
\pl{0}: \quad & e_0 \, \cap \, e_4 + a_1\,z - e_1,                                                 & (x, y, s, e_2, e_3) &= \underline{1}, \\
\pl{1}: \quad & e_1 \, \cap \, 1 + a_1\,x,                                                          & (y, z, s, e_3, e_4) &= \underline{1}, \\
\pl{2}: \quad & e_2 \, \cap \, e_3 + a_1\,x + a_{3,2}\,e_1,                                        & (y, z, s, e_0, e_4) &= \underline{1}, \\
\pl{3}: \quad & e_3 \, \cap \, a_1\,y\,s + a_{3,2}\,y\,e_4 - a_{2,1}\,s\,e_2 - a_{4,3}\,e_2\,e_4,  & (x, z, e_0, e_1)    &= \underline{1}, \\
\pl{4}: \quad & e_4 \, \cap \, a_1\,y - e_3 - a_{2,1}\,e_0,                                        & (x, z, s, e_1, e_2) &= \underline{1}. 
\end{align}
Their intersection structure is
\begin{equation}\label{surface:intersection}
\pl{0} - \pl{1} - \pl{2} - \pl{3} - \pl{4} \,(- \,\pl{0}),
\end{equation}
which is the intersection structure of the extended Dynkin diagram $\tilde{A}_4$, as expected for an $SU(5)$-singularity.\\

The connection between roots and $\pl{}$s is trivial in the generic case: Each $\pl{i}$ corresponds to $-\alpha_i$.\\

%%%%%%%%%%%%%%%%%%%%%%%%%%%%%%%%%%%%%%%%%%%%%%%%%%%%
\subsection{Enhancements Curves} \label{app_curves}

\subsubsection*{Representation $\mathbf{10}_1$ on the Curve $\mathbf{\{a_1\}}$}
On the locus $a_1 = 0$ several of the $\pl{}$s split and the following $\pl{}$s appear:
\begin{align}\label{10_1:p1s}
  &\pl{03}: && e_0 \, \cap \, e_3,                                                           & (x, y, z, s, e_2) &= \underline{1},     \\
  &\pl{0A}: && e_0 \, \cap \, e_4 - e_1,                                                     & (x, y, s, e_2) &= \underline{1},  \\
  &\pl{13}: && e_1 \, \cap \, e_3,                                                           & (x, y, z, s) &= \underline{1},       \\
  &\pl{14}: && e_1 \, \cap \, e_4,                                                          & (x, y, z, s) &= \underline{1},       \\
  &\pl{24}: && e_2 \, \cap \, e_4,                                                           & (x, y, z, s, e_0) &= \underline{1},  \\
  &\pl{2B}: && e_2 \, \cap \, e_3 + a_{3,2}\,e_1,                                         & (y, z, s, e_0) &= \underline{1},     \\
  &\pl{3C}: && e_3 \, \cap \, a_{3,2}\,y\,e_4 - a_{2,1}\,s\,e_2 - a_{4,3}\,e_1\,e_2\,e_4,    & (x, z, e_0) &= \underline{1},     \\
  &\pl{4D}: && e_4 \, \cap \, e_3 + a_{2,1}\,e_0,                                            & (x, z, s, e_2) &= \underline{1}.
\end{align}\\
Depending on the triangulation $T_{ij}, i=1,2, j=1,2,3$ as defined after (\ref{srideal_optional}) the splitting process becomes:
\begin{equation}\label{10_1:splitting}
 \begin{array}{c||c|c|c}
  \,\,\textrm{Original}\,\,      &  T_{i1}         &  T_{i2}             &  T_{i3}                       \\
  \hline
  \pl{0}   &  \pl{0A}                          &  \pl{0A}            &  \pl{03} + \pl{0A}            \\
  \pl{1}   &  \pl{14}                          &  \pl{13} + \pl{14}  &  \pl{13}                      \\
  \pl{2}   &  \pl{24} + \pl{2B}                &  \pl{2B}            &  \pl{2B}                      \\
  \pl{3}   &  \pl{3C}                          &  \pl{13} + \pl{3C}  &  \,\,\pl{03} + \pl{13} + \pl{3C}\,\,  \\
  \pl{4}   &  \,\,\pl{14} + \pl{24} + \pl{4D}\,\,  &  \,\,\pl{14} + \pl{4D}\,\,  &  \pl{4D}
 \end{array}
\end{equation}\\
The intersection structure is
\begin{equation}\label{10_1:intersections}
 \begin{split}
   \pl{0A}\\\pl{4D}
 \end{split}
 > \pl{03}/\pl{14} - \pl{13}/\pl{24} <
 \begin{split}
   \pl{2B}\\\pl{3C}
 \end{split} \quad .
\end{equation}
Note that $\pl{03}/\pl{14} - \pl{13}/\pl{24}$ is short notation for one of the following three cases: $\pl{03} - \pl{13}$, $\pl{14} - \pl{13}$, or $\pl{14} - \pl{24}$. In particular, $\pl{03}$ and $\pl{24}$ never occur in the same triangulation. Then for each triangulation the above gives the structure of the extended Dynkin diagram $\tilde{D}_5$ associated to $SO(10)$.\\

The root representation of each $\pl{}$ becomes, again depending on the triangulation:
\small
\begin{equation}\label{10_1:root_representation}
 \begin{array}{c||c|c|c}
   \pl{i}   &  \textrm{Root } (T_{i1})        &  \textmd{Root } (T_{i2})      &   \textrm{Root } (T_{i3})   \\
   \hline
   \pl{03}  &  \cdot                                       &  \cdot                                      &  \mu_{10} - \alpha_2 - \alpha_3  \\
   \pl{0A}  &  \alpha_1 + \alpha_2 + \alpha_3 + \alpha_4   &  \alpha_1 + \alpha_2 + \alpha_3 + \alpha_4  &  \,\,-\mu_{10} + \alpha_1 + 2\,\alpha_2 + 2\,\alpha_3 + \alpha_4\,\,  \\
   \pl{13}  &  \cdot                                       &  \mu_{10} - \alpha_1 - \alpha_2 - \alpha_3  &  -\alpha_1  \\
   \pl{14}  &  -\alpha_1                                   &  -\mu_{10} + \alpha_2 + \alpha_3            &  \cdot  \\
   \pl{24}  &  \,\,-\mu_{10} + \alpha_1 + \alpha_2 + \alpha_3\,\,  &  \cdot                                      &  \cdot  \\
   \,\,\pl{2B}\,\,  &  \mu_{10} - \alpha_1 - 2\alpha_2 - \alpha_3  &  -\alpha_2                                  &  -\alpha_2  \\
   \pl{3C}  &  -\alpha_3                                   &  -\mu_{10} + \alpha_1 + \alpha_2            &  -\mu_{10} + \alpha_1 + \alpha_2  \\
   \pl{4D}  &  \mu_{10} - \alpha_2 - \alpha_3 - \alpha_4   &  \,\,\mu_{10} - \alpha_2 - \alpha_3 - \alpha_4\,\,  &  -\alpha_4  
 \end{array}
\end{equation}\\
\normalsize

The $\pl{}$-combination for the various states of the $\mathbf{10}$-representation are:
\footnotesize
\begin{equation}\label{10_1:p1_combination}
 \begin{array}{c|c c c c c c c c}
  \textrm{Weight}                                           &  \multicolumn{8}{c}{\pl{i}-\textrm{Combination } (T_{i1}, T_{i2}, T_{i3})}  \\
  \hline
                                                          &  \pl{03}         & \pl{0A} & \pl{13}     & \pl{14}     & \pl{24}         & \pl{2B} & \pl{3C} & \pl{4D} \\
  \hline
  \hline
  0                                                       &  (\cdot,\cdot,2) & (1,1,1) & (\cdot,2,2) & (2,2,\cdot) & (2,\cdot,\cdot) & (1,1,1) & (1,1,1) & (1,1,1)  \\
  \hline
  \mu_{10}                                                &  (\cdot,\cdot,2) & (1,1,1) & (\cdot,1,1) & (1,1,\cdot) & (0,\cdot,\cdot) & (0,0,0) & (0,0,0) & (1,1,1)  \\
  \mu_{10} - \alpha_2                                     &  (\cdot,\cdot,2) & (1,1,1) & (\cdot,1,1) & (1,1,\cdot) & (1,\cdot,\cdot) & (1,1,1) & (0,0,0) & (1,1,1)  \\
  \mu_{10} - \alpha_1 - \alpha_2                          &  (\cdot,\cdot,2) & (1,1,1) & (\cdot,2,2) & (2,2,\cdot) & (1,\cdot,\cdot) & (1,1,1) & (0,0,0) & (1,1,1)  \\
  \mu_{10} - \alpha_2 - \alpha_3                          &  (\cdot,\cdot,1) & (1,1,0) & (\cdot,2,0) & (1,1,\cdot) & (1,\cdot,\cdot) & (1,1,0) & (1,1,0) & (1,1,0)  \\
  \mu_{10} - \alpha_1 - \alpha_2 - \alpha_3               &  (\cdot,\cdot,1) & (1,0,0) & (\cdot,1,1) & (2,0,\cdot) & (1,\cdot,\cdot) & (1,0,0) & (1,0,0) & (1,0,0)  \\
  \mu_{10} - \alpha_2 - \alpha_3 - \alpha_4               &  (\cdot,\cdot,1) & (0,0,0) & (\cdot,0,0) & (0,0,\cdot) & (0,\cdot,\cdot) & (0,0,0) & (0,0,0) & (1,1,1)  \\
  \mu_{10} - \alpha_1 - 2\alpha_2 - \alpha_3              &  (\cdot,\cdot,1) & (0,0,0) & (\cdot,1,1) & (0,0,\cdot) & (0,\cdot,\cdot) & (1,1,1) & (0,0,0) & (0,0,0)  \\
  \mu_{10} - \alpha_1 - \alpha_2 - \alpha_3 - \alpha_4    &  (\cdot,\cdot,1) & (0,0,0) & (\cdot,1,1) & (1,1,\cdot) & (0,\cdot,\cdot) & (0,0,0) & (0,0,0) & (1,1,1)  \\
  \mu_{10} - \alpha_1 - 2\alpha_2 - \alpha_3 - \alpha_4   &  (\cdot,\cdot,1) & (0,0,0) & (\cdot,1,1) & (1,1,\cdot) & (1,\cdot,\cdot) & (1,1,1) & (0,0,0) & (1,1,1)  \\
  \mu_{10} - \alpha_1 - 2\alpha_2 - 2\alpha_3 - \alpha_4  &  (\cdot,\cdot,2) & (0,0,0) & (\cdot,2,0) & (1,1,\cdot) & (1,\cdot,\cdot) & (1,1,1) & (1,1,1) & (1,1,1)  \\
 \end{array}
\end{equation}
\normalsize
Note that the overall torus has Cartan charge $0$ so the first line gives the multiplicities of the various $\pl{}$s for each triangulation.\\

%%%%%%%%%%%%%%%%%%%%%%%%%%
\subsubsection*{Representation $\mathbf{5}_{3}$ on the Curve $\mathbf{\{a_{3,2}\}}$}
On $a_{3,2} = 0$, the following additional $\bbP^1$s appear:
\begin{align}\label{5_3:p1s}
\pl{2s}: \quad & e_2 \, \cap \, s,            & (y, z, e_0, e_1, e_4) &= \underline{1},  \\
\pl{2E}: \quad & e_2 \, \cap \, e_3 + a_1\,x,         & (y, z, e_0, e_4) &= \underline{1},  \\
\pl{3x}: \quad & e_3 \, \cap \, x,            & (y, z, e_0, e_1, e_4) &= \underline{1},  \\
\pl{3F}: \quad & e_3 \, \cap \, a_1\,y\,s - a_{2,1}\,x\,s\,e_2 - a_{4,3}\,e_4\,e_2,           & (z, e_0, e_1) &= \underline{1}.
\end{align}
The splitting procedure becomes, depending on the triangulation:
\begin{equation}\label{5_3:splitting}
 \begin{array}{c||c|c}
 \qquad \textmd{Original}    \qquad   &  \qquad T_{1j}    \qquad       &  \qquad T_{2j}           \qquad         \\
  \hline
 \pl{2} &   \pl{2E}  & \pl{2s} + \pl{2E} \\
 \pl{3} &  \pl{3x} + \pl{3F}  &  \pl{3F} \\
\end{array}
\end{equation}
The intersection structure is
\begin{align}\label{5_3:intersection}
 \pl{0} - \pl{1} - \pl{2E} - \pl{2s}/\pl{3x} - \pl{3F} - \pl{4} \,(- \,\pl{0}) ,
\end{align}
which for each triangulation gives the structure of the extended Dynkin diagram $\tilde{A}_5$ associated with $SU(6)$.\\

The root representation of each $\pl{}$ becomes, again depending on the triangulation:
\begin{equation}\label{5_3:root_representation}
 \begin{array}{c||c|c}
  \qquad   \pl{i}    \qquad &  \qquad \textmd{Roots for }  T_{1j}  \qquad    &    \qquad \textmd{Roots for } T_{2j} \qquad    \\
  \hline
    \pl{2s}                 &   \cdot                                        &    \mu_5-\alpha_1-\alpha_2                     \\
    \pl{2E}                 &   -\alpha_2                                    &    -\mu_5+\alpha_1                             \\
    \pl{3x}                 &   -\mu_5+\alpha_1+\alpha_2                     &    \cdot                                       \\
    \pl{3F}                 &   \mu_5-\alpha_1-\alpha_2-\alpha_3             &    -\alpha_3                                   
 \end{array}
\end{equation}\\

The $\pl{}$-combination for the various states of the $\mathbf{5}$-representation are:
\small
\begin{equation}\label{5_3:p1_combination}
 \begin{array}{c|c c c c c c c}
 \textmd{ Weight}                                              &  \multicolumn{7}{c}{  \pl{i}\textmd{-Combination }  (T_{1j}, T_{2j})  }  \\
  \hline
  &  \quad \pl{0} \quad  & \quad  \pl{1} \quad  &  \quad \pl{2s}  \quad & \quad  \pl{2E} \quad  &  \quad \pl{3x} \quad  &  \quad \pl{3F} \quad  &  \quad \pl{4}  \quad   \\
  \hline
  \hline
    0                                                        &  (1,1)   &   (1,1)   &   (\cdot,1)   &   (1,1)   &   (1,\cdot)   &   (1,1)   &   (1,1)    \\
  \hline
    \mu_{5}                                                  &    (1,1)   &   (0,0)   &   (\cdot,1)   &   (0,0)   &   (0,\cdot)   &   (1,1)   &   (1,1)   \\
    \mu_{5} - \alpha_1                                       &    (1,1)   &   (1,1)   &   (\cdot,1)   &   (0,0)   &   (0,\cdot)   &   (1,1)   &   (1,1)   \\
    \mu_{5} - \alpha_1 - \alpha_2                            &    (1,0)   &   (1,0)   &   (\cdot,1)   &   (1,0)   &   (0,\cdot)   &   (1,0)   &   (1,0) \\
    \mu_{5}- \alpha_1 - \alpha_2 - \alpha_3                  &    (0,0)   &   (0,0)   &   (\cdot,1)   &   (0,0)   &   (0,\cdot)   &   (1,1)   &   (0,0)  \\
    \mu_{5} - \alpha_1 - \alpha_2 - \alpha_3  - \alpha_4     &    (0,0)   &   (0,0)   &   (\cdot,1)   &   (0,0)   &   (0,\cdot)   &   (1,1)   &   (1,1)    
 \end{array}
\end{equation}\\
\normalsize

%%%%%%%%%%%%%%%%%%%%%%%%%%
\subsubsection*{Representation $\mathbf{5}_{-2}$ on the Curve $\mathbf{\{a_{3,2} a_{2,1} - a_{1} a_{4,3}\}}$}
On $a_{3,2}a_{2,1} - a_{1} a_{4,3}$, the following additional $\bbP^1$s occur:
\begin{align}\label{5_-2:p1s}
 \pl{3G}: \quad & e_3 \, \cap \, a_1\,x + a_{3,2},    & (z, s, e_0, e_1, e_4) &= \underline{1}.  \\
 \pl{3H}: \quad & e_3 \, \cap \, a_1\,y - a_{2,1}\,s,      & (x, z, e_0, e_1, e_2) &= \underline{1}. 
\end{align}
The splitting process is simply the following:
\begin{equation}\label{5_-2:splitting}
 \begin{array}{c||c}
 \qquad \textmd{Original}    \qquad   &  \qquad T_{i j}    \qquad            \\
  \hline
 \pl{3} & \pl{3G} + \pl{3H}\\
\end{array}
\end{equation}
The intersection structure is
\begin{align}\label{5_-2:intersection}
 \pl{0} - \pl{1} - \pl{2} - \pl{3G} - \pl{3H} - \pl{4} \,(- \,\pl{0}),
\end{align}
which is the structure of the extended Dynkin diagram $\tilde{A}_5$ associated with $SU(6)$.\\

The root representation of each $\pl{}$ becomes
\begin{equation}\label{5_-2:root_representation}
 \begin{array}{c||c}
  \qquad   \pl{i}    \qquad &  \qquad \textmd{Roots for } T_{i j}  \qquad \\
  \hline
  \pl{3G}               &    -\mu_5 + \alpha_1 + \alpha_2            \\
  \pl{3H}               &    \mu_5-\alpha_1-\alpha_2-\alpha_3        
 \end{array}
\end{equation}\\

The $\pl{}$-combinations for the various states of the $\mathbf{5}$-representation are:
\small
\begin{equation}\label{5_-2:p1_combination}
 \begin{array}{c|c c c c c c}
 \textmd{ Weight}                                 &  \multicolumn{6}{c}{  \pl{i}\textmd{-Combination} }  \\
  \hline
&  \quad \pl{0} \quad  & \quad  \pl{1} \quad  &  \quad \pl{2}  \quad & \quad  \pl{3G} \quad  &  \quad \pl{3H} \quad  &  \quad \pl{4} \quad\\
  \hline
  \hline
    0                                                        &    1   &   1   &   1   &   1   &   1   &   1  \\
  \hline
    \mu_{5}                                                  &   1   &   0   &   0   &   0   &   1   &   1   \\
    \mu_{5} - \alpha_1                                       &   1   &   1   &   0   &   0   &   1   &   1   \\
    \mu_{5} - \alpha_1 - \alpha_2                            &   1   &   1   &   1   &   0   &   1   &   0   \\
    \mu_{5}- \alpha_1 - \alpha_2 - \alpha_3                  &   0   &   0   &   0   &   0   &   1   &   1   \\
    \mu_{5} - \alpha_1 - \alpha_2 - \alpha_3  - \alpha_4     &   0   &   0   &   0   &   0   &   1   &   0   
 \end{array}
\end{equation}\\
\normalsize

%%%%%%%%%%%%%%%%%%%%%%%%%%%%%%%%%%%%%%%%%%%%%%%%%%%%
\subsection{Enhancements Points}\label{app_points}
%%%%%%%%%%%%%%%%%%%%%%%%%%
\subsubsection*{Yukawa Coupling $\mathbf{10\,10\,5}$ on $\mathbf{a_1 \cap \, a_{2,1}}$}
On $a_1 = a_{2,1} = 0$, the following $\bbP^1$ appear in addition to those appearing on $a_1 = 0$
\begin{align}\label{10105:p1s}
\pl{34}: \quad & e_3 \, \cap \, e_4,                              & (x, z, s) &= \underline{1},  \\
\pl{3J}: \quad & e_3 \, \cap \, a_{3,2}\,y - a_{4,3} ,            & (x, z, e_0, e_1, e_2) &= \underline{1}. 
\end{align}
Now there are two splitting processes: One starting from the $SO(10)$-curve on $a_1=0$, the other starting from the $SU(5)$-curve on $a_{2,1} a_{3,2} - a_{1} a_{4,3}$.
For the first of these, the splitting is relatively simple:
\begin{equation}\label{10105:10splitting}
 \begin{array}{c||c|c|c}
 \quad \textmd{Original} \quad & \qquad T_{i 1} \qquad       &  \qquad T_{i 2}    \qquad   &  \qquad T_{i 3}    \qquad  \\
 \hline
 \pl{3C} & \pl{34} + \pl{3J} & \pl{34} + \pl{3J} & \pl{03} + \pl{34} + \pl{3J}\\
 \pl{4D} & \pl{24} + \pl{34} & \pl{34}               & \pl{34}
\end{array}
\end{equation}
with all other $\pl{}$s invariant. The second splitting process is slightly more involved and depends again on the triangulation. In particular, note that $\pl{3H}$ becomes trivial:
\begin{equation}\label{10105:5splitting}
 \begin{array}{c||c|c|c}
  \,\,\textrm{Original}\,\, &  T_{i1}                          &  T_{i2}                       &  T_{i3}                                    \\
  \hline
  \pl{0}   &  \pl{0A}                         &  \pl{0A}                      &  \pl{03} + \pl{0A}                         \\
  \pl{1}   &  \pl{14}                         &  \pl{13} + \pl{14}            &  \pl{13}                                   \\
  \pl{2}   &  \pl{24} + \pl{2B}               &  \pl{2B}                      &  \pl{2B}                                   \\
  \pl{3G}  &  \pl{34} + \pl{3J}               &  \,\,\pl{13} + \pl{34} + \pl{3J}\,\,  &  \,\,2\,\pl{03} + \pl{13} + \pl{34} + \pl{3J}\,\,  \\
  \pl{3H}  &  \cdot                           &  \cdot                        &  \cdot                                     \\
  \pl{4}   &  \,\,\pl{14} + 2\,\pl{24} + \pl{34}\,\,  &  \pl{14} + \pl{34}            &  \pl{34}
 \end{array}
\end{equation}
The intersection structure becomes
\begin{align}\label{10105:intersection}
 \pl{3J} - \pl{34} -< 
 \begin{split}
   \pl{03} / \pl{14} - \pl{0A}\\
   \pl{13} / \pl{24} - \pl{2B}\\
 \end{split} \quad ,
\end{align}
where, depending on the triangulation:
\begin{align*}
 \pl{34} -< \begin{split} \pl{03} / \pl{14} \\ \pl{13} / \pl{24} \end{split} = 
\begin{split} \underline{T_{i1}} \qquad \\           \pl{14} \\             |\, \\ \pl{34} - \pl{24} \end{split} \qquad \qquad
\begin{split} \underline{T_{i2}} \quad  \\           \pl{14} \\ \pl{34} - \,|\, \\           \pl{13} \end{split}
\begin{split} \underline{T_{i3}} \qquad \\ \pl{34} - \pl{03} \\             |\, \\           \pl{13} \end{split} \quad 
\end{align*}
These diagrams have the structure of $E_6, T_{3,3,3}, E_6$ respectively, i.e.\ they are not extended Dynkin diagrams and in particular not $\tilde{E}_6$.
This nicely reproduces the result of~\cite{Esole:2011sm}, however with the correct multiplicities ($i \in \{1, 2\}$).\\

Nevertheless, it is possible to express the $\pl{}$s in terms of the simple roots and the highest weights of the $\mathbf{5}$- and the $\mathbf{10}$ representations:
\small
\begin{equation}\label{10105:root_representation}
 \begin{array}{c||c|c|c}
   \pl{i}       &  \textrm{Root } (T_{i1})                         &  \textrm{Root } (T_{i2})                             &  \textrm{Root } (T_{i3})                    \\
   \hline
   \pl{03}  &  \cdot                                       &  \cdot                                      &  \mu_{10} - \alpha_2 - \alpha_3  \\
   \pl{0A}  &  \alpha_1 + \alpha_2 + \alpha_3 + \alpha_4   &  \alpha_1 + \alpha_2 + \alpha_3 + \alpha_4  &  \,\,-\mu_{10} + \alpha_1 + 2\,\alpha_2 + 2\,\alpha_3 + \alpha_4\,\,  \\
   \pl{13}  &  \cdot                                       &  \,\,\mu_{10} - \alpha_1 - \alpha_2 - \alpha_3\,\,  &  -\alpha_1  \\
   \pl{14}  &  -\alpha_1                                   &  -\mu_{10} + \alpha_2 + \alpha_3            &  \cdot  \\
   \,\,\pl{24}\,\,  &  \,\,-\mu_{10} + \alpha_1 + \alpha_2 + \alpha_3\,\,  &  \cdot                      &  \cdot  \\
   \pl{2B}  &  \mu_{10} - \alpha_1 - 2\alpha_2 - \alpha_3  &  -\alpha_2                                  &  -\alpha_2  \\
   \pl{34}  &  -\mu_{5} + \alpha_1 + \alpha_2            &  \mu_{10} - \alpha_2 - \alpha_3 - \alpha_4 &  -\alpha_4  \\
   \pl{3J}  &  \mu_{5} - \alpha_1 - \alpha_2 - \alpha_3  &  \mu_{5} - \alpha_1 - \alpha_2 - \alpha_3  &  \mu_{5} - \alpha_1 - \alpha_2 - \alpha_3 
 \end{array}
\end{equation}\\
\normalsize

The $\pl{}$-combination for the various roots and highest weights take the following form:
\small
\begin{equation}\label{10105:p1_combination}
 \begin{array}{c|c c c c c c c c}
  \textrm{Root}         &  \multicolumn{8}{c}{\pl{i}\textrm{-Combination } (T_{i1}, T_{i2}, T_{i3})}  \\
  \hline
               &  \pl{03}         & \pl{0A} & \pl{13}     & \pl{14}     & \pl{24}         & \pl{2B} & \pl{34} & \pl{3J} \\
  \hline
  \hline
  0              &  \,(\cdot,\cdot,3)\, & (1,1,1) & (\cdot,2,2) & (2,2,\cdot) & (3,\cdot,\cdot) & (1,1,1) & (2,2,2) & (1,1,1)  \\
  \hline
  \mu_{10}       &  (\cdot,\cdot,1) & \,(1,1,0)\, & (\cdot,1,1) & (1,1,\cdot) & (1,\cdot,\cdot) & (0,0,1) & (1,1,1) & (0,0,1)  \\
  \mu_{5}        &  (\cdot,\cdot,1) & (1,1,1) & \,(\cdot,0,0)\, & (1,1,\cdot) & (2,\cdot,\cdot) & (0,0,0) & (1,1,1) & (1,1,1)  \\
  \,-\alpha_0\,  &  (\cdot,\cdot,1) & (1,1,1) & (\cdot,0,0) & \,(0,0,\cdot)\, & (0,\cdot,\cdot) & (0,0,0) & (0,0,0) & (0,0,0)  \\
  -\alpha_1      &  (\cdot,\cdot,0) & (0,0,0) & (\cdot,1,1) & (1,1,\cdot) & \,(0,\cdot,\cdot)\, & (0,0,0) & (0,0,0) & (0,0,0)  \\
  -\alpha_2      &  (\cdot,\cdot,0) & (0,0,0) & (\cdot,2,2) & (1,1,\cdot) & (1,\cdot,\cdot) & \,(1,1,1)\, & (0,0,0) & (0,0,0)  \\
  -\alpha_3      &  (\cdot,\cdot,2) & (0,0,0) & (\cdot,1,1) & (0,0,\cdot) & (0,\cdot,\cdot) & (0,0,0) & \,(1,1,1)\, & (1,1,1)  \\
  -\alpha_4      &  (\cdot,\cdot,0) & (0,0,0) & (\cdot,0,0) & (1,1,\cdot) & (2,\cdot,\cdot) & (0,0,0) & (1,1,1) & \,(0,0,0)\,  
 \end{array}
\end{equation}\\
\normalsize

%%%%%%%%%%%%%%%%%%%%%%%%%%
\subsubsection*{Yukawa Coupling $\mathbf{10\,5\,5}$ on $\mathbf{a_1 \cap \, a_{3,2} }$}
On $a_1 = a_{3,2} = 0$, the only $\bbP^1$ occurring that has not appeared before is
\begin{align}\label{1055:p1s}
\pl{3K}: \quad & e_3 \cap a_{2,1}\,s + a_{4,3}\,e_4,		& (x, z, e_0, e_1) &= \underline{1}. 
\end{align}
The splitting process becomes
\small
\begin{equation}\label{1055:splitting}
 \begin{array}{c||c|c|c}
 \,\textrm{Original}\,  &  T_{11}   &  T_{12}    &  T_{13}  \\
 \hline
 \pl{0}  &  \pl{0A}  &  \pl{0A}  &  \pl{03} + \pl{0A}  \\
 \pl{1}  &  \pl{14}  &  \pl{13} + \pl{14}  &  \pl{13}  \\
 \pl{2}  &  \pl{23} + \pl{24}  &  \pl{23}  &  \pl{23}  \\
 \pl{3}  &  \,\,\pl{23} + \pl{3x} + \pl{3K}\,\,  &  \,\,\pl{13} + \pl{23} + \pl{3x} + \pl{3K}\,\,  &  \,\,\pl{03} + \pl{13} + \pl{23} + \pl{3x} + \pl{3K}\,\,  \\
 \pl{4}  &  \pl{14} + \pl{24} + \pl{4D}  &  \pl{14} + \pl{4D}  &  \pl{4D} \\[6pt]
 \hline
 \hline
 \textrm{Original}  &  T_{21} & T_{22}  &  T_{23} \\
 \hline
 \pl{0}  &  \pl{0A}  &  \pl{0A}  &  \pl{03} + \pl{0A}  \\
 \pl{1}  &  \pl{14}  &  \pl{13} + \pl{14}  &  \pl{13}  \\
 \pl{2}  &  \pl{23} + \pl{24} + \pl{2s}  &  \pl{23} + \pl{2s}  &  \pl{23} + \pl{2s}  \\
 \pl{3}  &  \pl{23} + \pl{3K}  &  \pl{13} + \pl{23} + \pl{3K}  &  \pl{03} + \pl{13} + \pl{23} + \pl{3K}  \\
 \pl{4}  &  \pl{14} + \pl{24} + \pl{4D}  &  \pl{14} + \pl{4D}  &  \pl{4D}
 \end{array}
\end{equation}
\normalsize
The intersection structure is
\begin{equation}\label{1055:intersection}
 \begin{split}
   \pl{0A}\\\pl{4D}
 \end{split}
 > \pl{03}/\pl{14} - \pl{13}/\pl{24} - \pl{23} <
 \begin{split}
   &\pl{2s}/\pl{3x}\\&\pl{3K}
 \end{split} \quad,
\end{equation}
which for each triangulation gives the structure of the extended Dynkin diagram $\tilde{D}_6$ associated with $SO(12)$.\\

The root representation of each $\pl{}$ becomes, again depending on the triangulation:
\small
\begin{equation}\label{1055:root_representation}
 \begin{array}{c||c|c|c}
   \pl{i}   &  \textrm{Root } (T_{i1})                     &  \textrm{Root } (T_{i2})                    &  \textrm{Root } (T_{i3})                      \\
   \hline
   \pl{03}  &  \cdot                                       &  \cdot                                      &  \mu_{10} - \alpha_2 - \alpha_3               \\
   \pl{0A}  &  \alpha_1 + \alpha_2 + \alpha_3 + \alpha_4   &  \alpha_1 + \alpha_2 + \alpha_3 + \alpha_4  &  -\mu_{10} + \alpha_1 + 2\,\alpha_2 + 2\,\alpha_3 + \alpha_4  \\
   \pl{13}  &  \cdot                                       &  \mu_{10} - \alpha_1 - \alpha_2 - \alpha_3  &  -\alpha_1                                    \\
   \pl{14}  &  -\alpha_1                                   &  -\mu_{10} + \alpha_2 + \alpha_3            &  \cdot                                        \\
   \pl{24}  &  -\mu_{10} + \alpha_1 + \alpha_2 + \alpha_3  &  \cdot                                      &  \cdot                                        \\
   \pl{3K}  &  -\mu_{5} + \alpha_1 + \alpha_2              &  -\mu_{5} + \alpha_1 + \alpha_2             &  -\mu_{5} + \alpha_1 + \alpha_2               \\
   \pl{4D}  &  \mu_{10} - \alpha_2 - \alpha_3 - \alpha_4   &  \mu_{10} - \alpha_2 - \alpha_3 - \alpha_4  &  -\alpha_4                                    \\[6pt]
   \hline
   \hline
   \pl{i}   &  \textrm{Root } (T_{11})                     &  \textrm{Root }(T_{12})                     &  \textrm{Root }(T_{13})  \\
   \hline
   \pl{23}  &  \mu_{10} - \alpha_1 - 2\alpha_2 - \alpha_3  &  -\alpha_2                                  &  -\alpha_2                                    \\
   \pl{3x}  &  -\mu_{5} + \alpha_1 + \alpha_2              &  -\mu_{5} + \alpha_1 + \alpha_2             &  -\mu_{5} + \alpha_1 + \alpha_2               \\[6pt]
   \hline
   \hline
   \pl{i}   &  \textrm{Root } (T_{21})                     &  \textrm{Root }(T_{22})                     &  \textrm{Root }(T_{23})  \\
   \hline
   \pl{23}  &  \mu_{5} - \alpha_1 - \alpha_2 - \alpha_3    &  -\mu_{5} + \alpha_1                        &  -\mu_{5} + \alpha_1                          \\
   \pl{2s}  &  \mu_{5} - \alpha_1 - \alpha_2               &  \mu_{5} - \alpha_1 - \alpha_2              &  \mu_{5} - \alpha_1 - \alpha_2                
 \end{array}
\end{equation}\\
\normalsize

The $\pl{}$-combination for the various roots and highest weights take the following form, where the order of triangulations is $(T_{11}, T_{12}, T_{13}, T_{21}, T_{22}, T_{23})$:
\small
\begin{equation}\label{1055:p1_combination}
 \begin{array}{c||c|c|c|c}
             &  0              &  \mu_{10}       &  \mu_{5\,(1)}     &  \mu_{5\,(2)}                            \\
  \hline
  \,\,\pl{03}\,\,  &  (0,0,2,\,\cdot,\,\cdot,2)  &  (\,\cdot,\,\cdot,2,\,\cdot,\,\cdot,2)  &  (\,\cdot,\,\cdot,2, \,\cdot,\,\cdot,2)  &  (\,\cdot,\,\cdot,2, \,\cdot,\,\cdot,2)  \\
  \pl{0A}  &  \,\,(1,1,1,1,1,1)\,\,  &  \,\,(1,1,1,1,1,1)\,\,  &  \,\,(1,1,1, 1,1,1)\,\,  &  \,\,(1,1,1, 1,1,1)\,\,  \\
  \pl{13}  &  (\,\cdot,2,2,\,\cdot,2,2)  &  (\,\cdot,1,1,\,\cdot,1,1)  &  (\,\cdot,1,1, \,\cdot,1,1)  &  (\,\cdot,1,1, \,\cdot,1,1)  \\
  \pl{14}  &  (2,2,\,\cdot,2,2,\,\cdot)  &  (1,1,0,1,1,0)  &  (1,1,0, 1,1,0)  &  (1,1,0, 1,1,0)  \\
  \pl{24}  &  (2,\,\cdot,\,\cdot,2,\,\cdot,\,\cdot)  &  (0,\,\cdot,\,\cdot,0,\,\cdot,\,\cdot)  &  (1,\,\cdot,\,\cdot, 1,\,\cdot,\,\cdot)  &  (1,\,\cdot,\,\cdot, 1,\,\cdot,\,\cdot)  \\
  \pl{23}  &  (2,2,2,2,2,2)  &  (0,0,0,0,0,0)  &  (1,1,1, 1,1,1)  &  (1,1,1, 1,1,1)  \\
  \pl{2s}  &  (\,\cdot,\,\cdot,\,\cdot,1,1,1)  &  (\,\cdot,\,\cdot,\,\cdot,0,0,0)  &  (\,\cdot,\,\cdot,\,\cdot, 1,1,1)  &  (\,\cdot,\,\cdot,\,\cdot,0,0,0)  \\
  \pl{3x}  &  (1,1,1,\,\cdot,\,\cdot,\,\cdot)  &  (0,0,0,\,\cdot,\,\cdot,\,\cdot)  &  (1,1,1, \,\cdot,\,\cdot,\,\cdot)  &  (0,0,0,\,\cdot,\,\cdot,\,\cdot)  \\
  \pl{3K}  &  (1,1,1,1,1,1)  &  (0,0,0,0,0,0)  &  (0,0,0, 1,1,1)  &  (1,1,1, 0,0,0)  \\
  \pl{4D}  &  (1,1,1,1,1,1)  &  (1,1,1,1,1,1)  &  (1,1,1, 1,1,1)  &  (1,1,1, 1,1,1)  \\[6pt]
  \hline
  \hline
           &  -\alpha_1      &  -\alpha_2      &  -\alpha_3      &  -\alpha_4  \\
  \hline
  \pl{03}  &  (\,\cdot,\,\cdot,0,\,\cdot,\,\cdot,0)  &  (\,\cdot,\,\cdot,0,\,\cdot,\,\cdot,0)  &  (\,\cdot,\,\cdot,0,\,\cdot,\,\cdot,0)  &  (\,\cdot,\,\cdot,0,\,\cdot,\,\cdot,0)  \\
  \pl{0A}  &  (0,0,0,0,0,0)  &  (0,0,0,0,0,0)  &  (0,0,0,0,0,0)  &  (0,0,0,0,0,0)  \\
  \pl{13}  &  (\,\cdot,1,1,\,\cdot,1,1)  &  (\,\cdot,0,0,\,\cdot,0,0)  &  (\,\cdot,1,1,\,\cdot,1,1)  &  (\,\cdot,0,0,\,\cdot,0,0)  \\
  \pl{14}  &  (1,1,\,\cdot,1,1,\,\cdot)  &  (0,0,\,\cdot,0,0,\,\cdot)  &  (0,0,\,\cdot,0,0,\,\cdot)  &  (1,1,\,\cdot,1,1,\,\cdot)  \\
  \pl{24}  &  (0,\,\cdot,\,\cdot,0,\,\cdot,\,\cdot)  &  (1,\,\cdot,\,\cdot,1,\,\cdot,\,\cdot)  &  (0,\,\cdot,\,\cdot,0,\,\cdot,\,\cdot)  &  (1,\,\cdot,\,\cdot,1,\,\cdot,\,\cdot)  \\
  \pl{23}  &  (0,0,0,0,0,0)  &  (1,1,1,1,1,1)  &  (1,1,1,1,1,1)  &  (0,0,0,0,0,0)  \\
  \pl{2s}  &  (\,\cdot,\,\cdot,\,\cdot,0,0,0)  &  (\,\cdot,\,\cdot,\,\cdot,1,1,1)  &  (\,\cdot,\,\cdot,\,\cdot,0,0,0)  &  (\,\cdot,\,\cdot,\,\cdot,0,0,0)  \\
  \pl{3x}  &  (0,0,0,\,\cdot,\,\cdot,\,\cdot)  &  (0,0,0,\,\cdot,\,\cdot,\,\cdot)  &  (1,1,1,\,\cdot,\,\cdot,\,\cdot)  &  (0,0,0,\,\cdot,\,\cdot,\,\cdot)  \\
  \pl{3K}  &  (0,0,0,0,0,0)  &  (0,0,0,0,0,0)  &  (1,1,1,1,1,1)  &  (0,0,0,0,0,0)  \\
  \pl{4D}  &  (0,0,0,0,0,0)  &  (0,0,0,0,0,0)  &  (0,0,0,0,0,0)  &  (1,1,1,1,1,1)  
 \end{array}
\end{equation}\\
\normalsize

%%%%%%%%%%%%%%%%%%%%%%%%%%
\subsubsection*{Yukawa Coupling $\mathbf{5\,5\,1}$ on $\mathbf{a_{3,2} \cap \, a_{4,3}}$}
Starting from the $\mathbf{5}_{3}$-locus $a_3 = 0$, only one additional $\pl{}$ appears:
\begin{align}\label{551:p1s}
\pl{3L}: \quad & e_3 \, \cap \, a_1\,y - a_{2,1}             & (x, z, e_0, e_1, e_2) &= \underline{1}
\end{align}
The splitting procedure becomes:
\begin{equation}\label{551:splitting}
 \begin{array}{c||c}
  \qquad \textmd{Original}    \qquad   &  \qquad T_{ij} \qquad  \\
  \hline
  \pl{3F} &  \pl{3s} \cdot \pl{3L} \\
\end{array}
\end{equation}
The intersection structure is:
\begin{equation}\label{551:intersection}
 \pl{0} - \pl{1} - \pl{2E} -  \pl{2s}/\pl{3x} - \pl{3s} - \pl{3L} - \pl{4} \,(- \,\pl{0}) ,
\end{equation}
which for each triangulation gives the structure of the extended Dynkin diagram $\tilde{A}_6$ associated to $SU(7)$.\\

The root representation of each $\pl{}$ becomes, depending on the triangulation:
\small
\begin{equation}\label{551:root_representation}
 \begin{array}{c||c|c}
  \qquad   \pl{i}    \qquad &  \qquad \textmd{Roots for }  T_{1j}  \qquad  &     \qquad    \textmd{Roots for } T_{2j}    \qquad                     \\
  \hline
    \pl{2E}                 &   -\alpha_2                                    &    -\mu_5+\alpha_1              \\
    \pl{2s}                 &   \cdot                                        &     \mu_5-\alpha_1-\alpha_2     \\
    \pl{3x}                 &   -\mu_5+\alpha_1+\alpha_2                     &    \cdot                        \\
    \pl{3s}                 &   \mu_1                                   &    -\mu_5+\alpha_1+\alpha_2          \\
    \pl{3L}                 &   \mu_5-\alpha_1-\alpha_2-\alpha_3             &    -\alpha_3                                    
 \end{array}
\end{equation}\\
\normalsize

The $\pl{}$-combination for the various roots and highest weights take the following form:
\small
\begin{equation}\label{551:p1_combination}
 \begin{array}{c|c c c c c c c}
  \,\textrm{Root}\,  &  \multicolumn{7}{c}{\pl{i}-\textrm{Combination } (T_{1j}, T_{2j})}  \\
  \hline
                 &  \pl{0}  &  \pl{1}  &  \pl{2E}  &  \pl{2s}  &  \pl{3x}  &  \pl{3F}  &  \pl{4}  \\[2pt]
  \hline
  \hline
    0            &  \,(1,1)\,   &  (1,1)   &   (1,1)   & (\cdot,1) & (1,\cdot) &   (1,1)   &   (1,1)  \\
  \hline
    \mu_{5}      &  (1,1)   &  \,(0,0)\,   &   (0,0)   & (\cdot,1) & (0,\cdot) &   (1,1)   &   (1,1)  \\
    \mu_{1}      &  (0,0)   &  (0,0)   &   \,(0,0)\,   & (\cdot,1) & (0,\cdot) &   (1,1)   &   (0,0)  \\
    \alpha_0     &  (1,1)   &  (0,0)   &   (0,0)   & \,(\cdot,0)\, & (0,\cdot) &   (0,0)   &   (0,0)  \\
    \alpha_1     &  (0,0)   &  (1,1)   &   (0,0)   & (\cdot,0) & \,(0,\cdot)\, &   (0,0)   &   (0,0)  \\
    \alpha_2     &  (0,0)   &  (0,0)   &   (1,1)   & (\cdot,1) & (0,\cdot) &   \,(0,0)\,   &   (0,0)  \\
    \alpha_3     &  (0,0)   &  (0,0)   &   (0,0)   & (\cdot,0) & (1,\cdot) &   (1,1)   &   \,(0,0)\,  \\
    \alpha_4     &  (0,0)   &  (0,0)   &   (0,0)   & (\cdot,0) & (0,\cdot) &   (0,0)   &   (1,1)  \\
 \end{array}
\end{equation}
\normalsize
\ 

%%%%%%%%%%%%%%%%%%%%%%%%%%%%%%%%%%%%%%%%%%%%%%%%%%%%%%%%%%%%%%%%%%%%%%%%
\subsection{Generic Structure on \texorpdfstring{$C_{34}$}{C34}}\label{sec:GenStructureOnC34}
Over the curve $C_{34}$, the Tate polynomial splits again and consequently the $\pl{}$s over this curve are given by $a_{3,2} \, \cap \, a_{4,3} \, \cap \, y_a$ intersected with one of the following two equations
\begin{align}
  &s = 0,  & (z = 1),\\
  &y^2\,e_3\,e_4 + a_1\,x\,y\,z = x^3\,s\,e_1\,e_2^2\,e_3 + a_{2,1}\,x^2\,z^2\,e_0\,e_1\,e_2.
\end{align}\\
Their intersection structure is
\begin{equation}
 \pl{A} == \pl{B},
\end{equation}
which is the extended Dynkin diagram $\tilde{A}_1$, associated to $SU(2)$.\\

%%%%%%%%%%%%%%%%%%%%%%%%%%%%%%%%%%%%%%%%%%%%%%%%%%%%%%%%%%%%%%%%%%%%%%%%
\subsection{Differences to the \texorpdfstring{$\pl{}$}{P1}-fibre structure for non-restricted \texorpdfstring{$SU(5)$}{SU(5)}-models}
\label{app_non-restricted}
In this section we summarise the differences of the $\pl{}$-fibre structure of non-restricted models to the one described above. In general the $\pl{}$s are very similar as only those change for which an additional $a_{6,5}$-term appears or for which removing $s$ has any relevance. Further, the elements of the Stanley-Reisner ideal $\{xy,\,zs\}$ are replaced by $\{xyz\}$, which may have an effect on which variables have to be nonzero in the partially inhomogeneous form used above.

In particular, in the fibre over generic points along the $SU(5)$ locus as well as over enhancement curves there is only one $\pl{}$ which changes in each case.\footnote{Of course for all $\pl{}$s, $s$ is removed from the list of variables that are set to $1$, if it is present.} In the following we list those $\pl{}$s that change their structure, restricting ourselves to the fibre over co-dimension one and -two singular loci,
\begin{align*}
  &\pl{3} \;\,\, \rightarrow  && e_3 \, \cap \, a_1\,y\,x + a_{3,2}\,y\,e_4 - a_{2,1}\,x^2\,e_2 - a_{4,3}\,x\,e_2\,e_4 - a_{6,5}\,e_1^2\,e_2\,e_4^2,  & (z, e_0, e_1)    &= \underline{1}, \\
  &\pl{3C} \rightarrow && e_3 \, \cap \, a_{3,2}\,y\,e_4 - a_{2,1}\,x^2\,e_2 - a_{4,3}\,x\,e_1\,e_2\,e_4 - a_{6,5}\,e_1^2\,e_2\,e_4^2,    & (z, e_0) &= \underline{1},     \\
  &\pl{3H} \rightarrow && e_3 \, \cap \, a_1\,a_{3,2}\,y - a_{2,1}\,a_{3,2}\,x - a_1\,a_{6,5}\,e_4,      & (z, e_0, e_1, e_2) &= \underline{1}.
\end{align*}
Here $\pl{3}$ is the relevant $\mathbb P^1$  over generic points along the $SU(5)$ curve and $\pl{3C}$ occurs over the ${\bf 10}$-curve $a_1=0$. Finally, $\pl{3H}$ lies, in the $U(1)$-restricted mode, over the locus $a_{2,1} a_{3,2} - a_1 a_{4,3}$, which, in the non-restricted case, takes the more general form $a_{3,2} \left(a_{2,1} a_{3,2} - a_1 a_{4,3}\right) + a_1^2 a_{6,5}$.\\

For the analysis of the recombined $\mathbf{5}$-curve it is also convenient to consider the intersection of $a_{6,5}$ with the $\mathbf{5}$-curve. This intersection splits into the two loci $a_{6,5} \cap a_{3,2}$ and $a_{6,5} \cap a_{2,1} a_{3,2} - a_1 a_{4,3}$. It is then of interest to consider the splitting of $\pl{3}$ above these loci. Above $a_{6,5} \cap a_{3,2}$ one obtains
\begin{align*}
 &\pl{3x} \rightarrow && e_3 \, \cap \, x,            & (z, e_0, e_1, e_4) &= \underline{1},  \\
 &\pl{3F} \rightarrow && e_3 \, \cap \, a_1\,y - a_{2,1}\,x - a_{4,3}\,e_4,           & (z, e_0, e_1, e_2) &= \underline{1},
\end{align*}
while above $a_{6,5} \cap a_{2,1} a_{3,2} - a_1 a_{4,3}$ one obtains
\begin{align*}
  &\pl{3G} \rightarrow && e_3 \, \cap \, a_1\,x + a_{3,2},    & (z, e_0, e_1, e_4) &= \underline{1}.  \\
  &\pl{3H} \rightarrow && e_3 \, \cap \, a_1\,a_{3,2}\,y - a_{2,1}\,a_{3,2}\,x,      & (z, e_0, e_1, e_2) &= \underline{1}. 
\end{align*}

%%%%%%%%%%%%%%%%%%%%%%%%%%%%%%%%%%%%%%%%%%%%%%%%%%%%%%%%%%%%%%%%%%%%%%%%%%%%%%%%%%%%%%%%%%%%%%%%
\section{Intersection Properties}
\label{app_Intersections}

\subsection{List of intersection numbers}

Here we collect some useful intersection numbers involving the resolution divisors of the $SU(5) \times U(1)_X$ model:
\bea
\label{EE} \int_{\hat Y_4}  && D_a \wedge D_b \wedge E_i \wedge E_j = - C_{ij} \int_{B_3} D_a \wedge D_b \wedge W, \\
\label{DE} \int_{\hat Y_4}  && D_a \wedge D_b \wedge D_c \wedge E_i = 0, \\
\label{ZE} \int_{\hat Y_4}  && D_a \wedge D_b \wedge Z \wedge E_i =   \delta_{0i}  \int_{B_3} D_a \wedge D_b \wedge W    , \\
\label{Zc1}\int_{\hat Y_4} && D_a \wedge D_b \wedge D_c \wedge Z = \int_{B_3}  D_a \wedge D_b \wedge D_c, \\
\label{ZZ} \int_{\hat Y_4} && D_a \wedge D_b \wedge Z \wedge Z = - \int_{B_3}  D_a \wedge D_b \wedge c_1(B_3), \\
\label{DDDS} \int_{\hat Y_4} && D_a \wedge D_b \wedge D_c \wedge S = \int_{B_3}  D_a \wedge D_b \wedge D_c, \\
\label{ZS} \int_{\hat Y_4} && D_a \wedge D_b \wedge Z \wedge S = 0, \\
\label{SE} \sum_i \int_{\hat Y_4}  && D_a \wedge D_b \wedge S \wedge E_i =  \int_{B_3}  D_a \wedge D_b \wedge W, \\
\label{SE04} \int_{\hat Y_4} && D_a \wedge D_b \wedge S \wedge E_1 = \int_{\hat Y_4} D_a \wedge D_b \wedge S \wedge E_4 = 0, \\
\label{SS} \int_{\hat Y_4} && D_a \wedge D_b \wedge S \wedge S = - \int_{B_3}  D_a \wedge D_b \wedge c_1(B_3).
\eea

Eq. (\ref{EE}) is the standard implementation of the intersection structure of the Dynkin diagram in the resolution divisors $E_i$ of a non-abelian singularity, with $C_{ij}$ the Cartan matrix for, in this case, $SU(5)$. Eq. (\ref{DE}) follows from the fact that the two-forms dual to the resolution divisors have only 'one leg along the fibre'.
The rationale for (\ref{ZE}) is the relation $E_0 + \sum_{i=1}^4 E_i= W$ with $W$ the pullback of the $SU(5)$ divisor in the base $B_3$, along with (\ref{DE}). This reflects the homological relation $\sum_{i=0}^1 {\mathbb P}^1_i = [T^2]$ for the resolution  ${\mathbb P}^1$ in the fibre, with $[T^2]$ the smooth fibre class. 
Eq. (\ref{ZZ}) is a consequence of $Z (Z + c_1(B_3)) = 0$ together with (\ref{Zc1}).
The relation (\ref{DDDS}) follows from the observation made after eq. (\ref{twXU1}) that the intersection of $S$ with any 3 divisor classes pulled back from $B_3$ equals their intersection with $Z$. The intersection of the sections $Z$ and $S$, however, vanishes, as indicated in (\ref{ZS}) since $\{zs \}$ is in the SR-ideal.
The intersections  (\ref{SE}) and (\ref{SE04}) follow from the considerations of the SR ideal, while the last relation, (\ref{SS}), follows from considerations analogous to those, which lead to (\ref{ZZ}) (see below).\\

\subsection{Derivation of Intersection Properties from the Stanley-Reisner Ideal}
The above relations can also be explicitly derived from the Stanley-Reisner ideal of the resolution manifold. From the optional elements
\begin{equation*}
  \left\{
  \begin{split}
   & x e_3 \\ & s e_2
  \end{split}
 \right\}
 \otimes
 \left\{
  \begin{split}
   & y e_0 \\ & z e_4
  \end{split}
 \right\}
 \otimes
 \left\{
  \begin{split}
   & x e_0,\, x e_1 \\ & x e_0,\, z e_2 \\ & z e_1,\, z e_2
  \end{split}
 \right\}
 \otimes
 \left\{
  \begin{split}
   & e_0 e_3,\, e_1 e_3 \\ & e_0 e_3,\, e_2 e_4 \\ & e_1 e_4,\, e_2 e_4
  \end{split}
 \right\}
\end{equation*}
one can choose $z e_4$ and $z e_1,\, z e_2$ from columns two and three - since on the Calabi-Yau manifold $Z$ never intersects any of the $E_i$,
the possibility of them intersecting in the ambient space is irrelevant to the analysis. Then these three properties along with the ones following from elements of the SR-ideal which appear for all triangulations can be used to
derive the above relations. First of all one obtains
\begin{equation}
 Z\,E_1 = Z\,E_2 = Z\,E_3 = Z\,E_4 = 0,\qquad S\,E_1 = S\,E_4 = 0.
\end{equation}
Furthermore one finds:
\begin{align*}
 s e_0: && S\,P          = &\, S\,(E_2 + E_3),                         \\
 y e_1: && 3\,c_1\,E_1\, = &\, (1,2,3,2)_k E_k\, E_1,                  \\
 y e_2: && \left(3\,c_1 - S\right)E_2\, = &\, (1,2,3,2)_k E_k\, E_2,   \\
 x e_4: && \left(2\,c_1 - S\right)E_4\, = &\, (1,2,2,1)_k E_k\, E_4,   \\
 x y  : && 6\left(Z + c_1\right)^2 - 5\,c_1\,S + S^2 = &\, -S\left(5\,P - 3\,E_2\right) + c_1 (2,4,12,3)_k E_k, \\
                                                    &&& - \left(E_1 + 2\,E_2\right)\left(2\,E_3-E_4\right) - 6\,E_3^{\,2} - 3\,E_2\,E_4.
\end{align*}

The first column of the above SR-ideal options leads to one of the following intersection properties:
\begin{align*}
  s e_2: && S\,E_2 =&\, 0 ,                                               \\
  x e_3: && \qquad \qquad \quad \left(2\,c_1 - S\right)E_3\, = &\, (1,2,2,1)_i E_i\, E_3. \qquad \qquad \qquad \qquad \qquad \quad \,\ \\
\end{align*}

In addition to these properties one trivially finds that the intersection of four base divisors classes is zero, as is the intersection of 
three base divisor classes with one of the exceptional classes. Combining these properties and noting that $[P_W] = 6(Z+c_1) - S - (2,4,5,3)_k E_k$, one finds
\begin{align*}
 [P_W] (Z - S) = & 6(Z+c_1)Z - 6\,c_1\,S + S^2 + S(5\,P-E_2) \\
             = & -6(Z+c_1)c_1 - c_1\,S + 2\,S\,E_2 + c_1 (2,4,12,3)_k E_k \\
               & - \left(E_1 + 2\,E_2\right)\left(2\,E_3-E_4\right) - 6\,E_3^{\,2} - 3\,E_2\,E_4
\end	{align*}
\begin{equation*}
 \Rightarrow \qquad \qquad \qquad \qquad \qquad [P_W] (Z - S) D_a\,D_b\,D_c = 0 \qquad \qquad \qquad \qquad \qquad 
\end{equation*}
showing that $S$ gives the same intersection numbers (with three base divisors) as the section defined by $[z=0]$.
This section adheres to the usual relation, present already in non-resolved, non-restricted models:
\begin{align*}
 xyz:  && (Z + c_1)^2 Z &= 0,\\
 \Rightarrow && \qquad \qquad \qquad \qquad [P_W]\,Z\,(Z + c_1) &= 0 .\qquad \qquad \qquad \qquad \qquad \qquad \qquad \ 
\end{align*}

To find a similar property for $S$, one first considers the term $xys$ and then substitutes this into the corresponding expression for $S$:
\begin{align*}
 &xys:        && S^2 [5\,c_1 - S - 5\,P + E_2] = S \left[6\,c_1^{\,2} - c_1 (12\,P - 2\,E_2) + 2\,P(P-E_2)\right].\\[10pt]
 &\Rightarrow && [P_W]\,S\,(S+c_1) = \,c_1\,S\left(6\,c_1 - 5\,P + E_2\right) + S^2\left(5\,c_1 - S - 5\,P + E_2\right).\\
 &            && \qquad \qquad \qquad \quad   = S\left[12\,c_1^{\,2} - 17\,P\,c_1 + 3\,E_2\,c_1 + 2\,P^2 - 2\,P\,E_2\right] .\\[10pt]
 &\Rightarrow && \qquad \qquad \qquad [P_W]\,S\,(S+c_1)\,D_a\,D_b = 0,
\end{align*}
where $D_a, D_b$ are proper transforms of base divisor classes.\\

%%%%%%%%%%%%%%%%%%%%%%%%%%%%%%%%%%%%%%%%%%%%%%%%%%%%%%%%%%%%%%%%
\section{Fibre Ambient Space}
\label{app_Ambient}
For the reader's convenience in this appendix we collect some aspects of the fibre ambient space that arises in the blow-up resolution process described in \ref{sec:resolution}. First of all, let us focus on the scaling relation induced by each blow-up. For the new blow-up coordinate $e_i^n$ these will be
\begin{equation}
 \begin{array}{c|c|c|c}
      x             &      y             &  \,\,w\,/\,e_0\,\,  &  \,\,e_i^n\,\,  \\
  \hline
  \,\,2\,n - a\,\,  &  \,\,3\,n - b\,\,  &      -n         &      1
 \end{array}
\end{equation}
where $(a, b)$ is an element of the following ordered list:
\begin{equation}
 \left[ (2,3),\, (1,2),\, (1,1),\, (0,1),\, (0,0),\, (-1,0),\, (0,-1) \right]_{\,i}
\end{equation}
which are the points of the toric diagram for $\mathbb{P}_{231}$. Note that $e_1^1 = w\,/\,e_0$. The sets of $e_i^n$ necessary to resolve a particular singularity above the GUT surface are encoded for various singularities in table 3.1 in~\cite{Candelas:1997eh}, part of which we reproduce in the following (the format being $n: i$ in each entry):
\begin{equation}
 \begin{array}{r|l l l}
  SU(5)  \quad  &  \quad 1: 2,3,4,5         &                &     \\
  SU(6)  \quad  &  \quad 1: 2,3,4,5,6 \quad &                &     \\
  SO(10) \quad  &  \quad 1: 3,4,5;          & 2: 1,2         &     \\
  SO(12) \quad  &  \quad 1: 3,5,6;          & 2: 1,2,4       &     \\
  E_6    \quad  &  \quad 1: 4,5;            & 2: 1,2,3;\quad & 3:1
 \end{array}
\end{equation}\\
 
%%%%%%%%%%%%%%%%%%%%%%%%%%%%%%%%%%%%%%%%%%%%%%%%%%%%%%%%%%%%%%%%%%

\bibliography{rev-flux}  
\bibliographystyle{utphys}
\end{document}